\documentclass[journal,doublecolumn]{IEEEtran}
\usepackage[T1]{fontenc}
\usepackage{color}
\usepackage{verbatim}
\usepackage{float}
\usepackage{booktabs}
\usepackage{bm}
\usepackage{amsmath}
\usepackage{amsthm}
\usepackage{amssymb}
\usepackage{caption}
\usepackage {graphicx}
\usepackage{subcaption}
\usepackage[outdir=./]{epstopdf}
\makeatletter

\providecommand{\tabularnewline}{\\}
\floatstyle{ruled}
\newfloat{algorithm}{tbp}{loa}
\providecommand{\algorithmname}{Algorithm}
\floatname{algorithm}{\protect\algorithmname}

\theoremstyle{plain}
\newtheorem{thm}{\protect\theoremname}
\theoremstyle{plain}
\newtheorem{prop}[thm]{\protect\propositionname}
\theoremstyle{remark}
\newtheorem{rem}[thm]{\protect\remarkname}
\theoremstyle{definition}
\newtheorem{defn}[thm]{\protect\definitionname}
\theoremstyle{plain}
\newtheorem{lem}[thm]{\protect\lemmaname}

\usepackage{cite}
\usepackage{caption}

\usepackage{babel}
\providecommand{\definitionname}{Definition}
\providecommand{\lemmaname}{Lemma}
\providecommand{\propositionname}{Proposition}
\providecommand{\remarkname}{Remark}
\providecommand{\theoremname}{Theorem}

\begin{document}

\title{Blind Data Detection in Massive MIMO via $\ell_{3}$-norm Maximization
over the Stiefel Manifold }

\author{{\normalsize{}Ye Xue, }\textit{\normalsize{}Graduate Student Member}\textit{\emph{\normalsize{},}}\textit{\normalsize{}
IEEE}{\normalsize{}, Yifei Shen, }\textit{\normalsize{}Graduate Student Member}\textit{\emph{\normalsize{},
}}\textit{\normalsize{}IEEE}{\normalsize{},}\textit{\normalsize{}
}{\normalsize{}Vincent Lau,}\textit{\normalsize{} Fellow}\textit{\emph{\normalsize{},
}}\textit{\normalsize{}IEEE}{\normalsize{}, Jun Zhang, }\textit{\normalsize{}Senior
Member}\textit{\emph{\normalsize{},}}\textit{\normalsize{} IEEE}{\normalsize{},
and Khaled B. Letaief, }\emph{\normalsize{}Fellow}{\normalsize{},
}\emph{\normalsize{}IEEE}{\normalsize{} \thanks {Y. Xue, Y. Shen, V. Lau and K. B. Letaief are with the Department of Electronic and Computer Engineering, Hong Kong University of Science and Technology, Hong Kong (E-mail: yxueaf, yshenaw, eeknlau, eekhaled@ust.hk). J. Zhang is with the Department of Electronic and Information Engineering, The Hong Kong Polytechnic University, Hong Kong (E-mail: jun-eie.zhang@polyu.edu.hk). (The corresponding author is Y. Xue)}}}
\maketitle
\begin{abstract}
Massive MIMO has been regarded as a key enabling technique for 5G
and beyond networks. Nevertheless, its performance is limited by the
large overhead needed to obtain the high-dimensional channel information.
To reduce the huge training overhead associated with conventional
pilot-aided designs, we propose a novel blind data detection method
by leveraging the channel sparsity and data concentration properties.
Specifically, we propose a novel $\ell_{3}$-norm-based formulation
to recover the data without channel estimation. We prove that the
global optimal solution to the proposed formulation can be made arbitrarily
close to the transmitted data up to a phase-permutation ambiguity.
We then propose an efficient parameter-free algorithm to solve the
$\ell_{3}$-norm problem and resolve the phase-permutation ambiguity.
We also derive the convergence rate in terms of key system parameters
such as the number of transmitters and receivers, the channel noise
power, and the channel sparsity level. Numerical experiments will
show that the proposed scheme has superior performance with low computational
complexity.
\end{abstract}


\section{Introduction}

Massive multiple-input multiple-output (MIMO) can significantly enhance
spectral efficiency and reduce interference in cellular networks,
and thus has been regarded a key enabler for 5G and beyond networks\cite{rusek2012scaling}.
However, the benefits heavily rely on accurate channel state information
(CSI), which represents a major challenge, especially for systems
with higher frequencies, e.g., millimeter wave (mmWave) massive MIMO
systems \cite{xhycover2017}. It is largely due to the faster temporal
variations and the large dimension of the channel matrix. A common
approach is to send sufficient pilot symbols for reliable channel
estimation, which then enables coherent data detection. However, with
the limited coherence time and large dimension of the channel matrix,
the pilot overhead will easily occupy too much radio resource. In
addition, due to the limited number of orthogonal pilot sequences
in multi-cell systems, pilot contamination will further jeopardize
the performance of coherent massive MIMO systems \cite{elijah2015comprehensive}.
To overcome this difficulty, blind data detection methods have been
proposed to recover the data from the received signal without training
pilots \cite{zheng2002communication,muquet2002subspace,ngo2012evd}.
However, due to the degradation of the detection accuracy, traditional blind scheme can achieve a degrees of freedom (DoF) of $K(1-\frac{K}{T})$
for a massive MIMO system with $K$ transmit antennas and $M$ receive
antennas in a rich scattering environment where $T$ is the coherence
time \cite{zheng2002communication}, which is the same as that achieved
by a coherent massive MIMO system taking account of the pilot overhead
\cite{1193803}.

There have been some attempts to exploit the channel sparsity to reduce the pilot overhead. Many experimental studies
have indicated the sparsity of massive MIMO channels in the angular
domain due to the limited number of scatterers\cite{samimi20163,zhou2007experimental}.
By exploiting the sparsity of such channels, compressed-sensing-based
channel estimation can reduce the pilot training\cite{6816089}, hence
achieving a DoF of $K(1-c\frac{K}{T})$, where $c$ depends on the
channel sparsity level \cite{zhang2017blind}. However,
the pilots for sparse channel estimation still lead to a DoF loss
of the order of $\mathcal{O}(K/T)$.

Recently, it has been shown that, with blind scheme, exploiting the
sparsity structure of massive MIMO channels can further improve the
performance \cite{zhang2017blind,liu2019super,mezghani2018blind,8114344}.
Particularly, Zhang et. al \cite{zhang2017blind} showed that under
some regularity conditions, blind detection can achieve a DoF arbitrarily
close to $K(1-\frac{1}{T})$ for sparse massive MIMO channels. Approximate
message passing (AMP)-based algorithms were proposed in \cite{zhang2017blind}
and \cite{liu2019super} for blind data detection by exploiting the
sparsity of the channel, which showed a superior performance. However,
AMP-based approaches generally rely on certain assumptions on the
probability density functions (PDF) of the channel and data, which
is unrealistic in practical systems. Moreover, AMP-based approaches
require an iterative message passing algorithm that leads to significant
complexity due to the slow convergence rate for the large problem size in massive MIMO systems.
In \cite{mezghani2018blind}, a subspace-based method, which decomposes
the covariance of the received signal into the subspace of the channel
and refines the decomposition by exploiting the channel sparsity,
was proposed in blind detection for massive MIMO. However, a long
sample sequence of the received signal is required to estimate the
covariance, which is applicable only with a very long channel coherence
time and very low mobility. In addition, all of the
aforementioned works fall short on the theoretical aspect, namely,
they fail to provide a theoretical analysis for the achievable performance.

In this paper, we propose a novel formulation for blind data detection
for sparse massive MIMO channels, supported by an efficient algorithm
which does not require the knowledge of the  PDF
of the data and the channel. In addition, the proposed scheme does
not contain tuning parameters, and can be implemented
efficiently with a fast convergence rate. The theoretical justification of the formulation and
the convergence rate of the algorithm are also developed. The main
contributions are summarized as follows:
\begin{itemize}
\item \textbf{Data Concentration Property:}\footnote{In this paper, the term \emph{concentration} represents the \emph{concentration
of measure }phenomenon. This phenomenon can be informally expressed
as\emph{ ``A random variable that depends in a Lipschitz way on many
independent variables is essentially constant''\cite{talagrand1996new}.}} In this paper, we exploit the statistical information of the data
transmitted in general communication systems for a novel formulation
of the blind detection problem. Specifically, we show that under mild
conditions, there is a\emph{ data concentration} phenomenon, which
enables a simple blind sparse recovery formulation of a massive MIMO
system over the Stiefel manifold. 
\item \textbf{Blind Detection via $\ell_{3}$-norm Maximization:} It was
shown in \cite{zhang2017blind} that the channel sparsity leads to
a fundamental performance gain for the degrees of freedom of massive
MIMO systems. Hence our proposed formulation leverages the sparsity
structure of massive MIMO channels. Traditionally, $\ell_{1}$-norm
is the most widely used formulation to induce sparsity \cite{candes2007sparsity}.
However, the non-smooth nature of the $\ell_{1}$-norm results in
low-convergence speed and high complexity \cite{qu2014finding}. In
this paper, inspired by the smoothness and the fact that a high-order
norm promotes sparsity \cite{Underl4,quqingl4,zhang2019structured,li2018global},
we consider $\ell_{3}$-norm maximization over the Stiefel manifold
for blind data recovery of massive MIMO systems with sparse channels.
Furthermore, we show that the global maximizer of the formulation
is arbitrarily close to the true data up to a phase and permutation
ambiguity for a sufficiently large number of antennas. 
\item \textbf{Efficient Algorithm and Convergence Rate Analysis:} By taking
advantage of the smooth property of the $\ell_{3}$-norm and the geometric
structure of the Stiefel manifold, a parameter-free algorithm with
low complexity is proposed for blind data recovery for massive MIMO
systems. We show that under mild conditions, the proposed algorithm
converges to the stationary point of the $\ell_{3}$-norm maximization
problem. \textcolor{black}{We further show the convergence rate in
terms of the key system parameters.}%
\end{itemize}
The rest of the paper is organized as follows. In Section \ref{sec:Signal-Model},
we present the system model, the sparse MIMO channel and  elaborate
the data concentration property. In Section \ref{sec:-based-Nonconvex-Problem},
we present the $\ell_{3}$-norm-based manifold optimization problem
as a new formulation for blind data detection problem. We further
discuss the structural properties of the optimal solution to the new
formulation in the noiseless and noisy cases. A fast and parameter-free
algorithms is proposed and the corresponding convergence analysis
are given in Section \ref{sec:Proposed-Parameter-free-Algorith}.
Numerical simulation results are provided in Section \ref{sec:Simulation-results}.
Finally, Section \ref{sec:Conclusion} summarizes the work.

\emph{Notations}: $\bm{X}^{-1}$ and $\bm{X}^{H}$
denote the inverse and conjugate transpose of matrix $\bm{X}$,
respectively. $null(\bm{X})$ represents the null space of $\bm{X}$.
$|{\bm{X}}|$ is used to take an element-wise abstract
value. $diag[\bm{x}]$ represents a diagonal matrix constructed
by using $\bm{x}$ as the diagonal elements. The $i$-th row vector
and $j$-th column vector in $\bm{X}$ are $\bm{X}_{i,:}$
and $\bm{X}_{;,j}$, respectively. $X_{i,j}$ represents the element
in the $i$-th row and $j$-th column of $\bm{X}$. $\odot$ denotes
the Hadamard product. $\lceil\cdot\rceil$ denotes the ceiling operator.
$\langle{\bm{X}},\bm{Y}\rangle$ is the general
inner product of $\bm{X}$ and $\bm{Y}$. Finally, $||\bm{X}||_{F}$,
$||\bm{X}||$ and $||\bm{X}||_{p}$ are respectively, the
Frobenius norm, spectral norm and induced $\ell_{p}$ norm of matrix
$\bm{X}$. 

\section{System Model\label{sec:Signal-Model}}

In this section, we introduce the system model of the considered massive
MIMO system, followed by the sparse channel model and the data concentration
property. 

\subsection{System Model}

\begin{figure}
\centering{}\includegraphics[scale=0.4]{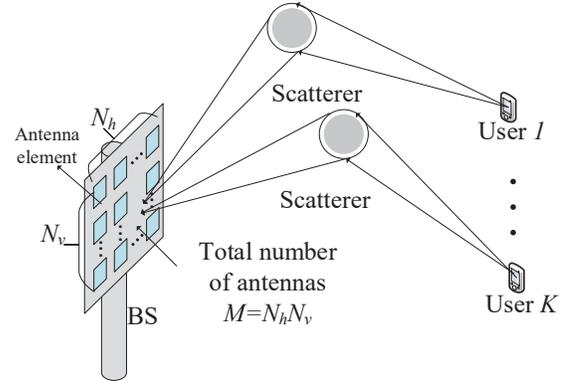}\caption{\label{fig:system} Illustration of the uplink of a single-cell multi-user
mmWave massive MIMO system. }
\end{figure}

Consider a single-cell mmWave massive MIMO communication system, as
illustrated in Fig. \ref{fig:system}. There are $K$ single-antenna
users transmitting to a base station (BS) that is equipped with $M$
receive antennas with $M\gg K\gg1$. Denote $\bm{X}_{k,:}\in\mathbb{C}^{1\times T}$
as the $T$ symbols transmitted by user $k\in\{1,\ldots K\}$ within
one frame. The aggregate transmit symbols of the $K$ users are denoted
by $\bm{X}\in\mathbb{C}^{K\times T}$. For simplicity, we consider
a block flat fading channel, but the framework can be easily extended
to OFDM systems. The received signal $\bm{Y}\in\mathbb{C}^{M\times T}$
at the BS is given by

\begin{equation}
\bm{Y}=\bm{H}\bm{G}^{1/2}\bm{P}^{1/2}\bm{X}+\bm{Z},\label{eq:sigmodel}
\end{equation}
where $\bm{P}=\text{diag}[P_{1,1},\ldots,P_{K,K}]$ is the transmit
power of the $K$ users, $\bm{H}=\text{[\ensuremath{\bm{H}_{:,1}},\ensuremath{\ldots},\ensuremath{\bm{H}_{:,K}}]}$
is the aggregate MIMO channel matrix, $\bm{H}_{:,k}\in\mathbb{C}^{M\times1}$
is the channel matrix between the $k$-th user and the BS, $\bm{G}=\text{diag}[G_{1,1}\ldots,G_{K,K}]$
is the aggregate large-scale fading coefficients of the $K$ users
to the BS, and $\bm{Z}\in\mathbb{C}^{M\times T}$  is the aggregate
additive channel noise of independent and identically distributed
(i.i.d.) $\mathcal{CN}(0,\sigma_{z}^{2})$ elements. We assume that
the BS has knowledge of $\bm{G}$, which can be obtained in practice
with a very low signaling overhead due to the slowly varying path
gain. 

\subsection{\label{subsec:Sparsity-of-the}Sparsity of the mmWave MIMO Channel }

The mmWave propagation environment is well characterized by a clustered
channel model \cite{el2014spatially}, which can be parameterized
by $N_{l}(k)$ paths of the $k$-th user. The small-scale mmWave channel
matrix $\bm{H}_{:,k}$ between user $k$ and the BS during the
coherence time is given by 

\begin{equation}
\bm{H}_{:,k}=\sqrt{\frac{M}{N_{l}(k)}}\sum_{l=1}^{N_{l}(k)}\alpha_{lk}\bm{a}_{r}(\varphi_{lk}^{r},\theta_{lk}^{r}),\label{eq:channelgen}
\end{equation}
where $\alpha_{lk}$ denotes the normalized path gain of the $l$-th
path for the $k$-th user. We assume that $\alpha_{lk}$ are i.i.d.
random variables following the complex Gaussian distribution $\mathcal{CN}(0,1)$,
$\varphi_{lk}^{r}$ and $\theta_{lk}^{r}$ denote the azimuth and
zenith angles of arrival (AoA) of the $l$-th path for the $k$-th
user, and $\bm{a}_{r}(\varphi_{lk}^{r},\theta_{lk}^{r})$ represents
the receive and transmit array response vectors. For simplicity, we
assume that the BS is equipped with a uniform rectangular planar array
(URPA) with $N_{h}$ and $N_{v}$ elements ($M=N_{h}N_{v}$) in the
horizontal and vertical direction, respectively. The array response
vector is given by \cite{balanis2016antenna} 

\begin{equation}
\begin{aligned}
&\bm{a}(\varphi,\theta)=\frac{1}{\sqrt{M}}[1,\ldots,e^{j\frac{2\pi}{\lambda}d(n_{v}sin(\varphi)sin(\theta)+n_{h}cos(\theta))},\ldots,\\
&e^{j\frac{2\pi}{\lambda}d((N_{v}-1)sin(\varphi)sin(\theta)+(N_{h}-1)cos(\theta))}]^{T},
\end{aligned}
\end{equation}
where $0\leq n_{v}\leq N_{v}$ and $0\leq n_{h}\leq N_{h}$.

\begin{figure}
	\centering
	\begin{subfigure}{.5\textwidth}
		\centering
		\includegraphics[width=.7\linewidth]{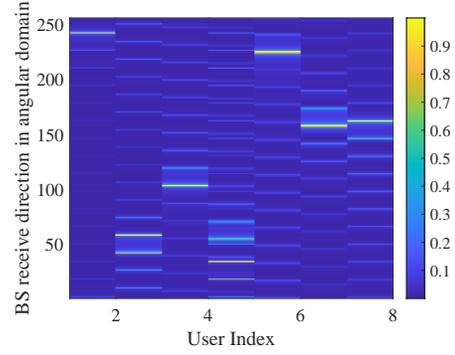}
		\caption{Simulated angular domain channel for $8$ single-antenna users.}
	\end{subfigure}%
\quad
\begin{subfigure}{.5\textwidth}
		\centering
		\includegraphics[width=.7\linewidth]{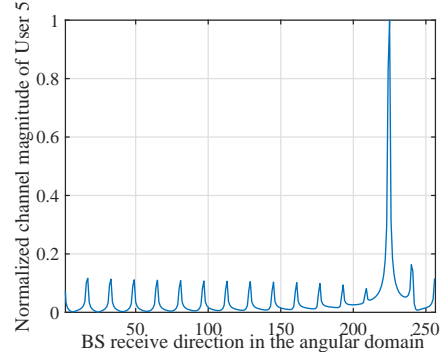}
		\caption{Simulated angular domain channel for user 5.}
	\end{subfigure}
	\caption{A simulated mmWave MIMO channel in the angular domain. The simulation
		uses the \emph{NYU WIRELESS 5G and 6G Millimeter Wave Statistical
			Channel Model }Matlab package \cite{samimi20163} under the $28$
		Ghz RMa NLoS drop-based model with a $16\times16$ UPA equipped at
		the receiver and single-antenna transmitter. $8$ independent trials
		are generated to mimic Eight independent users. The generated channel
		is then projected onto $\bm{U}_{M}$, as illustrated in (\ref{eq:angular}),
		to produce the angular domain channel $\bm{\bar{\bm{H}}}$.
		The pseudocolor plot of the magnitude of $\bm{\bar{\bm{H}}}$
		is shown in (a) and the magnitude of $\bm{\bar{\bm{H}}}_{:,5}$
		is given in (b). \label{fig:Simulated-mmWave-MIMO}}

\end{figure}

The spatial aggregate channel $\bm{H}=[\bm{H}_{:,1},\bm{H}_{:,2},\ldots,\bm{H}_{:,K}]$
of $K$ users can be expressed by a ``virtual angular domain'' representation
$\bar{\bm{H}}$ as follows:

\begin{equation}
\bm{H}=\bm{U}_{M}\bm{\bar{\bm{H}}},\label{eq:angular}
\end{equation}
where $\bm{U}_{M}$ is the steering matrix for the receive array.
With an $N_{h}\times N_{v}$ receive URPA, we have $\bm{U}_{M}=\bm{F}_{N_{v}}\otimes\bm{F}_{N_{h}}$\cite{brady2014beamspace},
where $\bm{F}_{N_{v}}\in\mathbb{C}^{N_{v}\times N_{v}}$ and $\bm{F}_{N_{h}}\in\mathbb{C}^{N_{h}\times N_{h}}$
are the unitary discrete Fourier transform (DFT) matrices. $\bar{H}_{m,k}$
can be interpreted as the normalized channel gain between the $k$-th
user and the $m$-th discrete receive angle \cite{sayeed2002deconstructing}.
From \cite{samimi20163}, the number of clusters is quite limited
in the mmWave band. Furthermore, $M$ is usually quite large to mitigate
the path loss effect in mmWave frequencies \cite{swindlehurst2014millimeter}.
Hence, the number of paths is usually much smaller than the channel
dimension, i.e., $N_{l}(k)\ll M$. Therefore, $\bm{\bar{\bm{H}}}$
can be regarded as \emph{approximately sparse (or
spiky)}.\footnote{The reason why $\bm{\bar{\bm{H}}}$ is not exactly sparse
is the mismatch between the exact receive angle and the predefined
one by $\bm{U}_{M}$, a.k.a. the energy leakage phenomenon.} In this paper, we define the sparsity level $\theta\in(0,1)$ as
the average number of non-zero elements in the sparse channel $\bm{\bar{\bm{H}}}$,
i.e., $\theta=\frac{||\bar{\bm{H}}||_{0}}{N_{h}N_{v}K}$.\emph{
}Fig. \ref{fig:Simulated-mmWave-MIMO} shows a realization of $\bar{\bm{H}}$
for a massive MIMO mmWave channel according to the statistical spatial
channel model (SSCM) implemented in NYUSIM \cite{samimi20163}. As
illustrated, most of the entries of $\bm{\bar{\bm{H}}}$ are
quite small (with $\theta\approx0.0132)$\footnote{Here we count a value that is less than 1\% of $||\bm{\bar{\bm{H}}}||_{\infty}$
as zero. } due to the limited number of scatterers in mmWave channels \cite{swindlehurst2014millimeter}. The channel sparsity can be further promoted through a more accurate
sparse basis $\mathbf{U}_{M}$, which can be found using the offline
learning method proposed in \cite{ding2018dictionary}.
 %

\subsection{Data Concentration on the Stiefel Manifold }

\begin{figure}[htpb]
\centering
\includegraphics[scale=0.5]{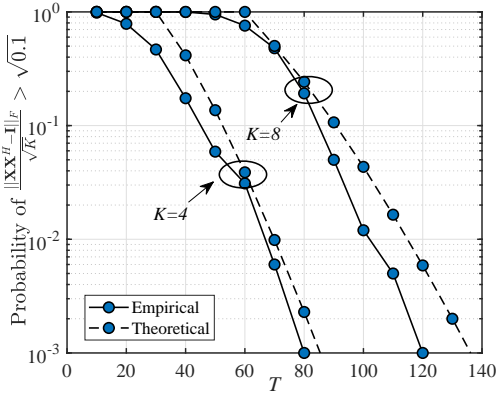}\caption{\label{fig:concentration1} Illustration of the data concentration.
The empirical curves are plotted by counting the frequency of $\frac{||\bm{X}\bm{X}^{H}-\bm{I}_{K}||_{F}}{\sqrt{K}}>\sqrt{0.1}$
over 1000 random trials. $\text{\ensuremath{\bm{X}}}$ is generated
with $T$ QPSK symbols from $K$ independent users. The $T$ QPSK
symbols are modulated from i.i.d $4T$ bits using Gray mapping and
normalized by $\frac{1}{\sqrt{T}}$, thus $\mathcal{S}_{\infty}=1$.
The theoretical curves are generated according to Proposition \ref{prop:Signal-concentration},
where the constant $C$ is chosen to be $0.416,0.464$ for $K=4,8$.}
\end{figure}

The aggregate data matrix transmitted by the $K$ users is assumed
to be independent with zero mean and a normalized covariance $\mathbb{E}[\bm{X}\bm{X}^{H}]=\bm{I}_{K}$.
The following proposition establishes that $\bm{X}\bm{X}^{H}$
is approaching $\bm{I}_{K}$ exponentially fast\footnote{One can use the simple Chebyshev's inequality to show that $P(||\bm{X}\bm{X}^{H}-\bm{I}_{K}||_{F}\geq\delta)<\frac{K^{4}Var(X^{2})}{T\delta^{2}}$,
where $Var(X^{2})$ is the variance of the square of each element
of $\bm{X}$, from the standard argument of the law of large numbers
\cite[Section 8.2, Theorem 2.1]{ross2006first}. However, our result
is stronger as it shows that the concentration is exponentially fast
with respect to $T$ under some mild conditions.} with respect to $T$. 
\begin{prop}
\label{prop:Signal-concentration}(Exponential concentration of data)
Assume the elements of matrix $\bm{X}\in\mathbb{C}^{K\times T}$
are independent, zero-mean and have bounded support given by $||\bm{X}||_{\infty}=\mathcal{S}_{\infty}/\sqrt{T}$
with $\mathbb{E}[\bm{X}\bm{X}^{H}]=\bm{I}_{K}$. Then,
there exists a constant $C\geq0$ such that, for any $\delta\geq0$,
we have\\
\begin{equation}
\begin{aligned}
&Pr\Big[\frac{||\bm{X}\bm{X}^{H}-\bm{I}_{K}||_{F}}{\sqrt{K}}\geq\frac{1}{\ln2}\mathcal{S}_{\infty}^{2}\max\{\delta,\delta^{2}\}\Big]\\
<&2\text{\ensuremath{\exp(}}-(\frac{\delta\sqrt{T}}{C}-\sqrt{K})^{2}),\quad\text{for }T\geq\frac{C^{2}K}{\delta^{2}},
\end{aligned}
\end{equation}

\end{prop}
\begin{IEEEproof}
See Appendix \ref{subsec:Proof-of-Proposition}.
\end{IEEEproof}
\begin{rem}
The conditions in Proposition \ref{prop:Signal-concentration} can
be satisfied by a general constellation in the transmit data of massive
MIMO systems. For example, the two types of constellations used in
5G systems, i.e., phase-shift keying (PSK) and quadrature amplitude
modulation (QAM), both have zero mean and bounded support. Specifically,
quadrature PSK (QPSK) symbols have $\mathcal{S}_{\infty}=1$ after
the power normalization. From Proposition \ref{prop:Signal-concentration},
as long as $T\geq\frac{(\sqrt{K}+\sqrt{\ln2})^{2}C^{2}}{\delta^{2}\ln^{2}2}$,
the probability of \emph{$\frac{||\bm{X}\bm{X}^{H}-\bm{I}_{K}||_{F}}{\sqrt{K}}>\delta$}
decays exponentially fast to $0$, as illustrated in Fig. \ref{fig:concentration1}. 
\end{rem}

Based on the data concentration property in Proposition \ref{prop:Signal-concentration},
we assume that $\bm{X}^{H}$ lies on a Stiefel manifold as defined
below.
\begin{defn}
\label{def:(Complex-Stiefel-manifold)}(Complex Stiefel manifold)
The complex Stiefel manifold $St_{K}(\mathbb{C}^{T})$ is defined
as the subspace of orthonormal $K$- frames in $\mathbb{C}^{T}$,
namely,

\begin{equation}
St_{K}(\mathbb{C}^{T})=\{\bm{\Gamma}\in\mathbb{C}^{T\times K}:\bm{\bm{\Gamma}}^{H}\bm{\Gamma}=\bm{I}_{K}\}.
\end{equation}
\end{defn}

\section{Problem Formulation: $\ell_{3}$-norm Maximization over the Stiefel
Manifold\label{sec:-based-Nonconvex-Problem}}

In this section, we formulate blind data recovery for massive MIMO
systems with sparse channels as an $\ell_{3}$-norm maximization problem
over the Stiefel manifold, and discuss the structural properties of
the optimal solution.

\subsection{Problem Formulation\label{subsec:Ideal-Noiseless-Case}}

\begin{figure}
\centering{}\includegraphics[scale=0.5]{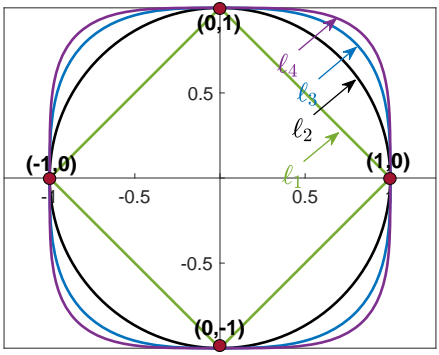}\caption{\label{fig:Unit-spheres-of}Unit spheres of the of $\ell_{p}$ in
$\mathbb{R}^{2}$, where $p=1,2,3,4$.}
\end{figure}

Following the discussion in Section \ref{subsec:Sparsity-of-the},
the received signal $\bm{Y}$ in (\ref{eq:sigmodel}) can be transformed
to
\begin{equation}
\bar{\bm{Y}}=\bm{U}_{M}^{H}\bm{Y}=\bm{\bar{\bm{H}}}\bm{G}^{1/2}\bm{P}^{1/2}\bm{X}+\bar{\bm{Z}},\label{eq:sigsparse}
\end{equation}
where $\bar{\bm{Z}}=\bm{U}_{M}^{H}\bm{Z}$. We first briefly
review some conventional approaches for blind data detection in massive
MIMO systems exploiting both the channel sparsity and the data concentration,
i.e., $\bm{X}\bm{X}^{H}\approx\bm{I}_{K}$. 
\begin{itemize}
\item \textbf{Conventional $\ell_{1}$-norm-based formulation}\cite{bai2018subgradient}

To exploit the data concentration property of $\bm{X}$, we observe
that $\bm{\bar{Y}}\bm{X}^{H}\bm{G}^{-1/2}\approx\bm{\bar{\bm{H}}}\bm{P}^{1/2}$
is also a sparse quantity with sparsity induced by $\bar{\bm{H}}$.
Hence, one may directly use the $\ell_{1}$-norm to promote sparsity
and formulate the blind data recovery as

\begin{equation}
\text{\ensuremath{\underset{\bm{A}\in St_{K}(\mathbb{C}^{T})}{\text{min}}}}\quad||\bar{\bm{Y}}\bm{A}\bm{G}^{-1/2}||_{1},\label{eq:Pl0-2}
\end{equation}
where the solution $\bm{A}^{*}$ is an estimate of $\bm{X}^{H}$.
However, Problem (\ref{eq:Pl0-2}) may lead to trivial solutions when
$T>K$. When $T>K$, $\bm{X}\in\mathbb{C}^{K\times T}$ will have
a null space with rank $T-K$. Hence $\bm{A}\in null(\bm{X})$
is a trivial solution to Problem (\ref{eq:Pl0-2}) with $||\bar{\bm{Y}}\bm{A}\bm{G}^{-1/2}||_{1}=0$.
When $T=K$, solving Problem (\ref{eq:Pl0-2}) will recover the data
$\bm{X}^{H}$. However, due to the non-smooth nature of the $\ell_{1}$-norm,
algorithms such as subgradient descent \cite{bai2018subgradient}
will have a high complexity. Furthermore, it was shown in \cite{Underl4}
that the formulation in (\ref{eq:Pl0-2}) is sensitive to noise because
the $\ell_{1}$-norm essentially encourages all small entries to be
0 \cite{zhang2019structured}.
\item \textbf{Complete dictionary learning approach}\cite{sun2016complete}

Another approach to exploit the property $\bm{X}\bm{X}^{H}\approx\bm{I}_{K}$
and the channel sparsity is to apply the complete dictionary method
\cite{sun2016complete}. Specifically, the approach recovers each
column of $\bm{X}^{H}$ sequentially\cite[Section 3]{sun2016complete2}.
To recover the $k$-th column of $\bm{X}^{H}$, we solve the following
problem: \\
\begin{equation}
\text{\ensuremath{\underset{||\bm{a}_{k}||_{2}=1}{\text{min}}}}\quad\frac{1}{M}\sum_{m=1}^{M}h_{\mu}(\bar{\bm{Y}}_{m,:}\bm{U}_{k-1}\bm{a}_{k}G_{k.k}^{-1/2}),\label{eq:col}
\end{equation}

where $h_{\mu}(\omega)=\mu\log\cosh(\omega/\mu)$ is to promote sparsity.
Consider $\bm{a}_{k}^{*}\in\mathbb{C}^{T-k+1}$ as the solution
of (\ref{eq:col}), then $\bm{U}_{k-1}\bm{a}_{k}^{*}\in\mathbb{C}^{T}$
is an estimation of $\bm{X}_{k,:}^{H}$, where $\bm{U}_{k-1}$
is an orthonormal basis for $[\text{span}(\bm{a}_{1}^{*},\bm{U}_{1}\bm{a}_{2}^{*},\ldots,\bm{U}_{k-2}\bm{a}_{k-1}^{*})]^{\perp}$.
However, this method requires that the dictionary $\bm{X}$ must
be a square matrix (complete)\cite{sun2016complete}. In massive MIMO
systems, we usually have a coherence time $T$ larger than the number
of users $K$. Hence, $\bm{X}\in\mathbb{C}^{K\times T}$ is usually
not square. Therefore, this formulation cannot be directly adopted
for the blind data recovery of massive MIMO systems.
\end{itemize}
As illustrated above, conventional approaches are either inefficient
or have strict requirements on the system parameters. Thus, an alternative
formulation to exploit both data concentration and channel sparsity
is needed. Recently, it has been shown in the machine learning literatures
\cite{zhang2019structured,Underl4} that maximizing a high-order norm
($\ell_{p}$-norm, $p>2$) leads to sparse (or spiky) solutions. An
intuitive explanation is that the sparsest points on the unit $\ell_{2}$-sphere,
e.g., points $(0,1)$, $(0,-1)$, $(1,0)$ and $(-1,0)$ in $\mathbb{R}^{2}$,
have the largest $\ell_{p}$-norm $(p>2)$, as shown in Fig. \ref{fig:Unit-spheres-of}. 

Inspired by this observation, we propose a smooth non-convex alternative
formulation of Problem (\ref{eq:Pl0-2}). The new formulation can
avoid the aforementioned issues caused by the conventional $\ell_{1}$-norm-based
formulation. It also relaxes the requirement $T=K$ on the coherence
time. The proposed problem is given by
\begin{equation}
\text{\ensuremath{\underset{\bm{A}\in St_{K}(\mathbb{C}^{T})}{\text{max}}}}\quad||\bar{\bm{Y}}\bm{A}\bm{G}^{-1/2}||_{3}^{3}.\label{eq:Pl3}
\end{equation}
In Problem (\ref{eq:Pl3}), we choose the cube of the $\ell_{3}$-norm
to promote sparsity, with the following justifications. First, maximization
of $||\bar{\bm{Y}}\bm{A}\bm{G}^{-1/2}||_{3}^{3}$ will
not lead to trivial solutions caused by minimizing the $\ell_{1}$-norm
formulation, i.e., $\bm{A}\in null(\bm{X})$, for $T>K$.
Second, the smoothness of the object function will make it possible
to design a fast-convergence algorithm. Third, $||\cdot||_{3}^{3}$
is a milder sparsity promoting function since it is very flat around
0, which will not encourage all small entries to be 0 and thus is
insensitive to small noise in the signal. Finally, $p=3$ can achieve
the smallest sample complexity\footnote{From a machine learning perspective, the sample complexity of a machine
learning algorithm represents the number of training-samples needed
to successfully learn a target function.} for exact recovery compared to other choices of $p$, $(p>2)$\cite{shen2020complete}.
In the following part, we shall provide theoretical support for formulation
(\ref{eq:Pl3}) by showing that solving it will recover the data matrix.
We start with the noiseless case, and then generalize the analysis
to the practical noisy case.

\subsection{Theoretical Analysis for Noiseless Case}

Without loss of generality, we assume $\bm{G=\bm{P}=\bm{I}}_{K}$
and $\bar{\bm{Z}}=\bm{0}$ for the noiseless case. We first
analyze the properties of Problem (\ref{eq:Pl3}) and show that the
global optimal solution is arbitrarily close to the true data $\bm{X}_{true}^{H}$
up to a \emph{phase-permutation ambiguity}. Specifically, the two
solutions, $\bm{A}_{1}$ and $\bm{A}_{2}$, are called \emph{equivalent
up to a phase permutation ambiguity} if $\bm{A}_{1}=\bm{\Xi}\bm{A}_{2}$,
where $\bm{\Xi}$ is a phase-permutation matrix as defined
below. 
\begin{defn}
\label{def:(Phase-permutation-Matrix)-The}(Phase-permutation matrix)
The $K$ dimensional phase-permutation matrix $\bm{\Xi}\in\mathbb{C}^{K\times K}$
is defined as:

\begin{equation}
\boldsymbol{\Xi}=\boldsymbol{\Sigma}\boldsymbol{\Pi},
\end{equation}
where $\boldsymbol{\Sigma}=diag(e^{j\phi_{1}},e^{j\phi_{2}},\ldots,e^{j\phi_{K}})$
with $\phi_{k}\in[0,2\pi]$ \emph{and $\boldsymbol{\Pi}=[\bm{e}_{\pi(1)},\bm{e}_{\pi(2)},\ldots,\bm{e}_{\pi(K)}]$,
with $\bm{e}_{k}$ being a standard basis vector, and $[\pi(1),\pi(2),\ldots,\pi(K)]$
being any permutations of the $K$ elements.}
\end{defn}
Note that a phase-permutation ambiguity can be resolved with very
small signaling overhead in massive MIMO systems. We shall defer the
discussion to Section \ref{subsec:Ambiguity-mitigation}. 

The following theorem summarizes the key result, namely, that the
optimal solution of Problem (\ref{eq:Pl3}) is arbitrarily close to
the true data $\bm{X}_{true}^{H}$ (up to a phase-permutation
ambiguity) for a sufficiently large number of antennas $M$, thus,
demonstrating the correctness of our formulation.
\begin{thm}
\textbf{\label{thm:Asymptotically-exact-recovery}}(Blind data detection
in the noiseless case) Let $\bm{\bm{\bar{\bm{H}}}}\in\mathbb{C}^{M\times K}$
with $\bar{H}_{m,k}\sim_{i.i.d}\mathcal{BG}(\theta)$\footnote{$\bar{H}_{m,k}$ is a product of an independent Bernoulli random variable
with parameter $\theta$ and a circular symmetric complex normal random
variable, i.e., $\bar{H}_{m,k}=g\odot b$, where $g\sim\mathcal{CN}(0,1)$,
$b\sim Ber(\theta)$. Theorem \ref{thm:Asymptotically-exact-recovery}
is based on the assumption that elements of $\bar{\bm{H}}\in\mathbb{C}^{M\times K}$
are i.i.d. Bernoulli-complex Gaussian random variables. The
i.i.d. assumption is only used for the convenience of analysis, and
the proposed blind scheme (from problem formulation to the algorithm) does not require this condition.}, $\bm{X}_{true}^{H}\in St_{K}(\mathbb{C}^{T})$, and $\bar{\bm{Y}}=\bm{\bar{\bm{H}}}\bm{X}_{true}$.
Define $\mathcal{A}^{*}$ as the set of optimal solutions of Problem
(\ref{eq:Pl3}) and $\bm{A}^{opt}\in\mathcal{A}^{*}$. For any
$\delta>0$, there exists a constant $c\geq0$, for $M\geq c\delta^{-2}K\log(K)(\theta K\log^{2}K)^{\frac{3}{2}}$,
such that 
\[
Pr\Big[\frac{1}{K}||\bm{A}^{opt}\boldsymbol{\Xi}-\bm{X}_{true}^{H}||_{F}^{2}\leq\delta\Big]\geq1-M^{-1},
\]
 where $\boldsymbol{\Xi}$ is a phase-permutation matrix.
\end{thm}
\begin{IEEEproof}
See Appendix \ref{subsec:Proof-of-Theorem}.
\end{IEEEproof}

\subsection{Theoretical Analysis for Noisy Case\label{subsec:Practical-Noisy-Case}}

In this section, we study the robustness of the solution to Problem
(\ref{eq:Pl3}) with respect to noise. The result is summarized in
the following theorem. 
\begin{thm}
\textbf{\label{thm:Asymptotically-exact-recovery-1}}(Blind data detection
in the noisy case) Let $\bm{\bm{\bar{\bm{H}}}}\in\mathbb{C}^{M\times K}$
with $\bar{H}_{m,k}\sim_{i.i.d}\mathcal{BG}(\theta)$, $\bm{X}_{true}^{H}\in St_{K}(\mathbb{C}^{T})$,
$\bm{P}=\bm{I}$, $\bar{\bm{Z}}\in\mathbb{C}^{M\times T}$
with $\bar{Z}_{m,t}\sim_{i.i.d}\mathcal{CN}(0,\sigma_{z}^{2})$, and
$\bar{\bm{Y}}=\bm{\bar{\bm{H}}}\bm{G}^{1/2}\bm{P}^{1/2}\bm{X}_{true}+\bar{\bm{Z}}$.
Define $\mathcal{A}^{*}$ as the set of optimal solutions of Problem
(\ref{eq:Pl3}) and $\bm{A}^{opt}\in\mathcal{A}^{*}$. For any
$\delta>0$, there exists a constant $c\geq0$, for $M\geq c\delta^{-2}K\log(K)(\theta K\log^{2}K)^{\frac{3}{2}}$,
such that 
\begin{equation}
\begin{aligned}
&Pr\Big[\frac{1}{K}||\bm{A}^{opt}\boldsymbol{\Xi}-\bm{X}_{true}^{H}||_{F}^{2}\leq\frac{\delta\xi(\sum_{k=1}^{K}(1+(G_{k,k}/\sigma_{z}^{2})^{-1}))^{\frac{3}{4}}}{K^{\frac{9}{4}}}\Big]\\
\geq&1-M^{-1},
\end{aligned}
\end{equation}

 where $\xi=\sum_{k=1}^{K}((1+(G_{k,k}/\sigma_{z}^{2})^{-1})^{\frac{3}{2}}+((G_{k,k}/\sigma_{z}^{2})^{-1})^{\frac{3}{2}})$
and $\boldsymbol{\Xi}$ is a phase-permutation matrix.
\end{thm}
\begin{IEEEproof}
See Appendix \ref{subsec:Proof-of-Proposition-1}.
\end{IEEEproof}
\begin{rem}
From Theorem \ref{thm:Asymptotically-exact-recovery-1}, there is
a high probability that the optimal solution to Problem (\ref{eq:Pl3})
will be within an \emph{uncertainty ball} centered at $\bm{X}_{true}^{H}$
(up to a phase-permutation ambiguity) with a radius of $\frac{\delta\xi(\sum_{k=1}^{K}(1+(G_{k,k}/\sigma_{z}^{2})^{-1})\text{)}^{\frac{3}{4}}}{K^{\frac{9}{4}}}$,
where $\sigma_{z}^{2}$ is the noise variance. The radius of the \emph{uncertainty
ball} decreases with decreasing $\sigma_{z}^{2}$ according to $\mathcal{O}((\sigma_{z}^{2})^{\frac{9}{4}})$.
When $\sigma_{z}^{2}\to0$, the result reduces to Theorem \ref{thm:Asymptotically-exact-recovery}
in the noiseless case.
\end{rem}

\section{A Low-complexity Parameter-free Algorithm\label{sec:Proposed-Parameter-free-Algorith}}

In this section, we propose a low-complexity parameter-free algorithm
to solve Problem (\ref{eq:Pl3}) and provide the corresponding convergence
analysis. The resolution of the phase-permutation ambiguity will also
be given.

\subsection{Review of the Gradient Method over the Stiefel Manifold}

Problem (\ref{eq:Pl3}) is an optimization problem over the Stefiel
manifold. As such, one can apply gradient search over the Stiefel
manifold \cite{absil2009optimization} to solve Problem (\ref{eq:Pl3}).
Specifically, the gradient iteration over the Stiefel manifold is
given by

\[
\bm{A}^{j+1}=Retr_{\bm{A}^{j}}(\tau^{j}grad\Psi(\bm{A}^{j})),
\]
where $j$ is the iteration index and $grad\Psi(\bm{A}^{j})$
denotes the gradient of the objective function $\Psi(\bm{A})$
at $\bm{A}^{j}\in St_{K}(\mathbb{C}^{T})$, which is given by
the orthogonal projection of the Euclidean gradient $\nabla\Psi(\bm{A}^{j})$
onto the tangent space at $\bm{A}^{j}$ \cite{absil2009optimization}.
$\tau^{j}$ is the stepsize to move in the direction $grad\Psi(\bm{A}^{j})$;
and $Retr_{\bm{A}}(\cdot)$ is the retraction on the manifold,
which maps $\tau^{j}grad\Psi(\bm{A}^{j})$ from the tangent space
onto the manifold itself. However, directly applying the gradient
method to Problem (\ref{eq:Pl3}) suffers from a high per iteration
complexity due to the two maps in each iteration. Specifically, to
compute $grad\Psi(\bm{A}^{j})$ one needs to projects the Euclidean
gradient $\nabla\Psi(\bm{A}^{j})$ onto the tangent space of $\bm{A}^{j}\in St_{K}(\mathbb{C}^{T})$
and $Retr_{\bm{A}^{j}}(\tau^{j}grad\Psi(\bm{A}^{j}))$ maps
the tangent vector $\tau^{j}grad\Psi(\bm{A}^{j})$ onto the Stiefel
manifold. Moreover, to find the optimal stepsize $\tau^{j}$, a curvilinear
search is required in each iteration, which is usually time-consuming.

\subsection{Parameter-free Algorithm to Solve Problem (\ref{eq:Pl3})}

To resolve the limitations of the gradient method, we propose a low-complexity
and parameter-free algorithm to solve Problem (\ref{eq:Pl3}). The
algorithm is derived based on the Frank-Wolfe method \cite{jaggi2013revisiting},
which considers a linear approximation of the objective function at
each iteration, hence substantially simplifying the per iteration
computation. 

\subsubsection{Derivation of the proposed parameter-free algorithm}

The Frank-Wolfe method for Problem (\ref{eq:Pl3}) iterates according
to

\begin{equation}
\begin{aligned}
\bm{S}^{j}:=\arg\max_{\bm{A}\in St_{K}(\mathbb{C}^{T})}\langle\nabla_{\bm{A}^{j}}||\bar{\bm{Y}}\bm{A}\bm{G}^{-1/2}||_{3}^{3},\bm{A}\rangle\\
{\color{blue}:=\arg\max_{\bm{A}\in Conv(St_{K}(\mathbb{C}^{T}))}\langle\nabla_{\bm{A}^{j}}||\bar{\bm{Y}}\bm{A}\bm{G}^{-1/2}||_{3}^{3},\bm{A}\rangle},\label{eq:fw1}
\end{aligned}
\end{equation}
{\color{blue} where $Conv(St_{K}(\mathbb{C}^{T}))$ is the convex hull of $St_{K}(\mathbb{C}^{T})$.}
\begin{align}
 & \bm{A}^{j+1}=(1-\upsilon^{j})\bm{A}^{j}+\upsilon^{j}\bm{S}^{j}\quad,\label{eq:upda}\\
\text{for\ensuremath{\quad}} & \upsilon^{j}\in\arg\max_{\upsilon\in[0,1]}||\bar{\bm{Y}}((1-\upsilon^{j})\bm{A}^{j}+\upsilon^{j}\bm{S}^{j})\bm{G}^{-1/2}||_{3}^{3}\nonumber 
\end{align}
where $\nabla_{\bm{A}^{j}}||\bar{\bm{Y}}\bm{A}\bm{G}^{-1/2}||_{3}^{3}$
is the Euclidean gradient at $\bm{A}^{j}$ of the objective function
of Problem (\ref{eq:Pl3}). We shall elaborate on these two steps
next.
\begin{itemize}
\item \textbf{Step 1: Simple computation of (\ref{eq:fw1}) by exploiting
the Stiefel manifold constraint.} We first focus on the computation
of (\ref{eq:fw1}). This step aims to find an $\bm{A}\in St_{K}(\mathbb{C}^{T})$
such that the inner product $\langle\nabla_{\bm{A}^{j}}||\bar{\bm{Y}}\bm{A}\bm{G}^{-1/2}||_{3}^{3},\bm{A}\rangle$
achieves the maximum, which can be obtained by projecting $\nabla_{\bm{A}^{j}}||\bar{\bm{Y}}\bm{A}\bm{G}^{-1/2}||_{3}^{3}$
onto the Stiefel manifold $St_{K}(\mathbb{C}^{T})$. This is similar
to \emph{retraction} in the gradient method, $Retr_{\bm{A}^{j}}(\tau^{j}grad\Psi(\bm{A}^{j}))$,
which maps the moving direction $\tau^{j}grad\Psi(\bm{A}^{j})$
onto the manifold. The difference is that \emph{retraction} is a mapping
from the tangent space of $\bm{A}^{j}$ to the manifold. However,
for the Stiefel manifold, polar-decomposition-based retraction, $Polar(\cdot)$,
is a projection-like retraction \cite{absil2012projection} which
does not require the direction to be in the tangent space. Therefore,
we can directly use polar-decomposition-based retraction to obtain\\
\begin{equation}
\begin{aligned}
&Polar(\nabla_{\bm{A}^{j}}||\bar{\bm{Y}}\bm{A}\bm{G}^{-1/2}||_{3}^{3})\\
:=&\arg\max_{\bm{A}\in St_{K}(\mathbb{C}^{T})}\langle\nabla_{\bm{A}^{j}}||\bar{\bm{Y}}\bm{A}\bm{G}^{-1/2}||_{3}^{3},\bm{A}\rangle.\label{eq:ploar}
\end{aligned}
\end{equation}
That is, $Polar(\nabla_{\bm{A}^{j}}||\bar{\bm{Y}}\bm{A}\bm{G}^{-1/2}||_{3}^{3})$
returns the matrix with orthonormal columns after the\emph{ right
polar decomposition}\cite{casey1983use} of matrix $\nabla_{\bm{A}^{j}}||\bar{\bm{Y}}\bm{A}\bm{G}^{-1/2}||_{3}^{3}$.
It turns out that the right polar decomposition has many fast computation
methods \cite{higham1994parallel}. In this paper, we use compact
singular value decomposition (SVD) to implement polar factorization.
By compact SVD, we have\\
\begin{equation}
\bm{U}\bm{\Sigma}\bm{V}^{H}=SVD_{compact}(\nabla_{\bm{A}^{j}}||\bar{\bm{Y}}\bm{A}\bm{G}^{-1/2}||_{3}^{3})\label{eq:svd}
\end{equation}
\end{itemize}
\begin{equation}
Polar(\nabla_{\bm{A}^{j}}||\bar{\bm{Y}}\bm{A}\bm{G}^{-1/2}||_{3}^{3})=\bm{U}\bm{V}^{H}.\label{eq:polar2}
\end{equation}

\begin{itemize}
\item \textbf{Step 2: Optimal step size in (\ref{eq:upda}) by exploiting
the convex objective function}. In general, the update Step (\ref{eq:upda})
requires a time-consuming line search. By exploiting the convexity
of the objective function of Problem (\ref{eq:Pl3}) as well as the
geometric structure of the Stiefel manifold, we show in Lemma \ref{lem:If--is}
that the optimal step size $\upsilon^{j}$ in (\ref{eq:upda}) is
given by $\upsilon^{j}=1$. 

\begin{lem}
\label{lem:If--is}(Optimal step size) $\upsilon^{j}=1$ is a solution
to $\max_{\upsilon\in[0,1]}||\bar{\bm{Y}}((1-\upsilon^{j})\bm{A}^{j}+\upsilon^{j}\bm{S}^{j})\bm{G}^{-1/2}||_{3}^{3}$.
\end{lem}
\begin{IEEEproof}
See Appendix \ref{subsec:Proof-of-Lemma-3}.
\end{IEEEproof}

\end{itemize}
Now, using (\ref{eq:ploar}) to solve (\ref{eq:fw1}) and by fixing
the step size as $\upsilon^{j}=1$ in (\ref{eq:upda}), we summarize
the main procedure of the proposed parameter-free algorithm to solve
Problem (\ref{eq:Pl3}) as 
\begin{equation}
\bm{A}^{j+1}=Polar(3\bar{\bm{Y}}^{H}(|\bar{\bm{Y}}\bm{A}^{j}\bm{G}^{-1/2}|\odot(\bar{\bm{Y}}\bm{A}^{j}\bm{G}^{-1/2}))\bm{G}^{-1/2}),\label{eq:itera}
\end{equation}
where $Polar(\cdot)$ can be calculated by (\ref{eq:svd}) and (\ref{eq:polar2}). 

\subsubsection{Convergence analysis}

Define\emph{ the first-order optimality metric} as 
\begin{equation}
\eta(\text{\ensuremath{\bm{A}}}^{j})=\underset{\bm{A}\in St_{K}(\mathbb{C}^{T})}{\max}\langle\bm{A}-\bm{A}^{j},\nabla_{\bm{A}^{j}}||\bar{\bm{Y}}\bm{A}\bm{G}^{-1/2}||_{3}^{3}\rangle.\label{eq:metricoptimality}
\end{equation}
The metric in (\ref{eq:metricoptimality}) is a measure of the optimality
of $\bm{A}^{j}$ due to the following lemma.\\

\begin{lem}
\label{lem:(Optimality-Measure)-}(Optimality Measure) $\text{\ensuremath{\bm{A}}}^{j}$
is a stationary point of Problem (\ref{eq:Pl3}), if and only if $\eta(\text{\ensuremath{\bm{A}}}^{j})=0$.
\end{lem}
\begin{IEEEproof}
See Appendix \ref{subsec: Proof-of-Lemmaopt}.
\end{IEEEproof}
Based on this, the following theorem summarizes the convergence of
the proposed algorithm to solve Problem (\ref{eq:Pl3}).
\begin{thm}
\label{lem:(Monotonically-increasing-value}(Convergence of the proposed
algorithm) Let $\{\bm{A}^{j}\}_{j=0}^{\infty}$ be the sequence
generated by the proposed algorithm in (\ref{eq:itera}) with a random
initial point $\bm{A}^{0}\in St_{K}(\mathbb{C}^{T})$. We have 
\end{thm}
\begin{enumerate}
\item $\{||\bar{\bm{Y}}\bm{A}^{j}\bm{G}^{-1/2}||_{3}^{3}\}_{j=0}^{\infty}$
is monotonically increasing.\label{enu:-is-monotonically}
\item $\underset{j\to\infty}{\lim}\eta(\text{\ensuremath{\bm{A}}}^{j})=0$.\label{enu:lim}
\item $\underset{0\leq i\leq j}{\min}\eta(\text{\ensuremath{\bm{A}}}^{i})\leq\frac{||\bar{\bm{Y}}\bm{A}^{opt}\bm{G}^{-1/2}||_{3}^{3}-||\bar{\bm{Y}}\bm{A}^{0}\bm{G}^{-1/2}||_{3}^{3}}{j+1}$.\label{enu:rate}
\end{enumerate}
\begin{IEEEproof}
See Appendix \ref{subsec:Proof-of-Theorem-dec}.
\end{IEEEproof}
\begin{rem}
Theorem \ref{lem:(Monotonically-increasing-value} shows that the
algorithm converges to a stationary point of Problem (\ref{eq:Pl3}),
with a rate $\mathcal{O}(1/j)$. 
\end{rem}
To evaluate the impact of the key system parameters on the convergence
rate of the proposed algorithm in (\ref{eq:itera}), we present the
following theorem. 
\begin{thm}
\label{thm:(Convergence-of-st-FW)-1}(Impact of the key system parameters)
Let $\{\bm{A}^{j}\}_{j=0}^{\infty}$ be the sequence generated
by the proposed algorithm in (\ref{eq:itera}) with a random initial
point $\bm{A}^{0}\in St_{K}(\mathbb{C}^{T})$. If the conditions
in Theorem \ref{thm:Asymptotically-exact-recovery-1} hold, for any
$\delta>0$, there exists a constant $c\geq0$, for $M\geq c\delta^{-2}K\log(K/\delta)(\sum_{k=1}^{K}(1+(G_{k,k}/\sigma_{z}^{2})^{-1})\theta\log K)^{\frac{3}{2}}\xi^{2}$,
such that,
\begin{equation}
\begin{aligned}
&Pr\Big[\underset{0\leq i\leq j}{\min}\eta(\text{\ensuremath{\bm{A}}}^{i})\\
&\leq(j+1)^{-1}(\frac{3}{4}\sqrt{\text{\ensuremath{\pi}}}M\sum_{k=1}^{K}\theta\Big(((G_{k,k}/\sigma_{z}^{2})^{-1}+1)^{\frac{3}{2}}\\
&-((G_{k,k}/\sigma_{z}^{2})^{-1})^{\frac{3}{2}}\Big)+2\delta)\Big]\geq1-M^{-1},\label{eq:rate}
\end{aligned}
\end{equation}
where $\xi=\sum_{k=1}^{K}((1+(G_{k,k}/\sigma_{z}^{2})^{-1})^{\frac{3}{2}}+((G_{k,k}/\sigma_{z}^{2})^{-1})^{\frac{3}{2}})$.
\end{thm}
\begin{IEEEproof}
See Appendix \ref{subsec:Proof-of-Theorem-noisy}.
\end{IEEEproof}
\begin{rem}
Theorem \ref{thm:(Convergence-of-st-FW)-1} shows that the convergence
rate is $\mathcal{O}(\frac{\theta M\sum_{k=1}^{K}\Big(((G_{k,k}/\sigma_{z}^{2})^{-1}+1)^{\frac{3}{2}}-((G_{k,k}/\sigma_{z}^{2})^{-1})^{\frac{3}{2}}\Big)}{j+1})$,
with high probability. This suggests that a smaller number of receive
antennas $M$, a smaller number of users $K$, a lower sparsity level
$\theta$, or a smaller noise variance $\sigma_{z}^{2}$ will lead
to a faster convergence rate. The convergence rate in (\ref{eq:rate})
is also consistent with the simulation results in Fig. \ref{fig:convespeed}.
\end{rem}

\subsubsection{Computational complexity analysis}
We analyze the computational complexity in terms
of the number of the floating-point operations (FLOPs). The main operations
of the proposed algorithm (\ref{eq:itera}) include: matrix multiplication,
 element-wise absolute value of a matrix,  element-wise product
of two matrices, and compact SVD of a matrix. The number of FLOPs
for the proposed algorithm  (\ref{eq:itera}) is
summarized in Table \ref{tab:Per-iteration-Complexity-Analysi}.

\begin{table}[htpb]
\centering\footnotesize{}\caption{\label{tab:Per-iteration-Complexity-Analysi}Per-iteration
Complexity Analysis}
\begin{tabular}{cc}
\toprule 
{\footnotesize{}Operation} & {\footnotesize{}Number of FLOPs\cite{hunger2005floating}}\tabularnewline
\midrule
\midrule 
{\footnotesize{}$\bm{P}_{1}=\bar{\bm{Y}}\bm{A}^{j}\bm{G}^{-1/2}$} & {\footnotesize{}$2MTK$}\tabularnewline
\midrule 
{\footnotesize{}$\bm{P}_{2}=|\bm{P}_{1}|$ }& {\footnotesize{}$MK$}\tabularnewline
\midrule 
{\footnotesize{}$\bm{P}_{3}=\bm{P}_{2}\odot\bm{P}_{1}$} & {\footnotesize{}$MK$}\tabularnewline
\midrule 
{\footnotesize{}$\bm{P}_{4}=\bar{\bm{Y}}^{H}\bm{P}_{3}\bm{G}^{-1/2}$} & {\footnotesize{}$2MTK$}\tabularnewline
\midrule 
{\footnotesize{}$[\mathbf{U},\sim,\mathbf{V}^{H}]=SVD_{compact}(\bm{P}_{4})$}&{\footnotesize{}$TK^{2}$\cite{alameddin2019toward}}\tabularnewline
\midrule 
{\footnotesize{}$\bm{A}^{j+1}=\bm{U}\bm{V}^{H}$} & {\footnotesize{}$TK^{2}$}\tabularnewline
\midrule 
{\footnotesize{}Total} & {\footnotesize{}$J\cdot\mathcal{O}(4MTK+2MK+2TK^{2})$}\tabularnewline
\bottomrule
\end{tabular}
\end{table}
{The complexity of the proposed method scales linearly
with both $M$ and $T$ and scales quadratically with $K$. In addition,
the proposed method does not require any tuning parameter, avoiding
the time-consuming parameter search. More importantly, the proposed
method enjoys a fast convergence rate as shown in Section \ref{sec:Simulation-results}}.

\subsection{Resolving Phase-permutation Ambiguity\label{subsec:Ambiguity-mitigation}}

\begin{figure}
	\centering
	\begin{subfigure}{.5\textwidth}
		\centering
		\includegraphics[width=.7\linewidth]{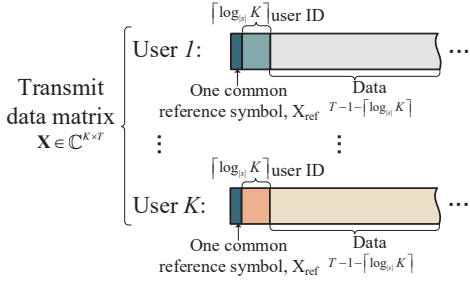}
		\caption{Illustration of the transmit data matrix, $\bm{X}$.}
		\label{fig:frameX}
	\end{subfigure}%
\quad
	\begin{subfigure}{.5\textwidth}
		\centering
		\includegraphics[width=.8\linewidth]{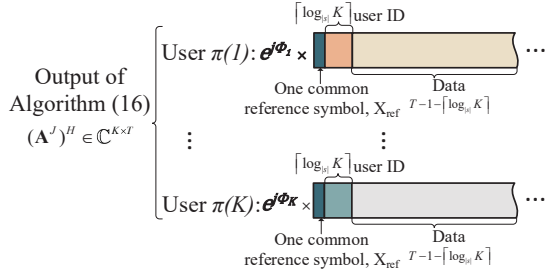}
		\caption{Illustration of the conjugate transpose of the
			output of Algorithm (\ref{eq:itera}) after $J$ iterations, $(\bm{A}^{J})^{H}$.}
			\label{fig:frameA}
	\end{subfigure}
	\caption{Illustration of the phase-permutation ambiguity resolution.\label{fig:Illustration-of-the}}
\end{figure}

According to Theorem \ref{thm:Asymptotically-exact-recovery-1}, there
is a phase-permutation ambiguity $\bm{\Xi}=\boldsymbol{\Sigma}\boldsymbol{\Pi}$
in the solutions of  Problem (\ref{eq:Pl3}). This ambiguity can be
resolved with little overhead \cite{liu2019super,zhang2017blind}.
Specifically, after we obtain the output of Algorithm (\ref{eq:itera}),
$(\bm{A}^{J})^{H}\in\mathbb{C}^{K\times T}$, the phase-permutation
ambiguity can be resolved by the following two steps \cite{liu2019super}:
\begin{itemize}
\item \textbf{Step 1: Using one common reference symbol to eliminate the
phase ambiguity.} Without loss of generality, we assume that the first
transmitted symbol of each user is the reference symbol, i.e., $X_{k,1}$=$X_{ref},\forall k\in\{1,\ldots K\}$
(as illustrated in Fig. \ref{fig:frameX}). The result after eliminating
the phase ambiguity is given by
\begin{equation}
\tilde{\bm{X}}=diag\Big(\frac{X_{ref}|(A^{J})_{1,1}^{H}|}{|X_{ref}|(A^{J})_{1,1}^{H}},\ldots,\frac{X_{ref}|(A^{J})_{K,1}^{H}|}{|X_{ref}|(A^{J})_{K,1}^{H}}\Big)(\bm{A}^{J})^{H},\label{eq:phase}
\end{equation}
where $diag(\frac{X_{ref}|(A^{J})_{1,1}^{H}|}{|X_{ref}|(A^{J})_{1,1}^{H}},\ldots,\frac{X_{ref}|(A^{J})_{K,1}^{H}|}{|X_{ref}|(A^{J})_{K,1}^{H}})$
is an estimation of $\boldsymbol{\Sigma}^{-1}$, as illustrated in
Fig. \ref{fig:frameA}.
\item \textbf{Step 2: Using user ID to eliminate the permutation ambiguity.
}After eliminating the phase ambiguity, the permutation ambiguity
can be eliminated by comparing the user ID with $\lceil\log_{|\mathcal{S}|}K\rceil$
symbols, where $|\mathcal{S}|$ is the size of the modulation alphabet
$\mathcal{S}$,as illustrated in Fig. \ref{fig:Illustration-of-the}.
\end{itemize}

Table \ref{tab:overhead} compares overheads for different schemes,
which, together with the achievable rate comparison in Fig. \ref{fig:mmwavec}
shows that the proposed blind detection scheme achieves a higher achievable
rate by saving the pilot overhead.

\begin{table}
\centering{}{\caption{ Overhead comparison\label{tab:overhead}}
}%
\begin{tabular}{cc}
\hline 
{Scheme} & {Overhead}\tabularnewline
\hline 
\hline 
{Pilot-based coherent detection} & {$K$}\tabularnewline
\hline 
{Proposed blind data detection} & {1+$\lceil\log_{|\mathcal{S}|}K\rceil$}\tabularnewline
\hline 
\end{tabular}
\end{table}

The overall algorithm is summarized by Algorithm \ref{alg2}.

\begin{algorithm}[H]
{\caption{\label{alg2}Proposed parameter-free algorithm}
}

{Input: $\bm{\bar{Y}}$}

{Output: $\bm{\hat{X}}$}

{Initialize: random $\bm{A}^{0}\in St_{K}(\mathbb{C}^{T})$.}

{for $j=0,\ldots,J$ do}

{1: $\bm{U}\bm{\Sigma}\bm{V}^{H}=SVD_{compact}(3\bar{\bm{Y}}^{H}(|\bar{\bm{Y}}\bm{A}^{j}\bm{G}^{-1/2}|\odot(\bar{\bm{Y}}\bm{A}^{j}\bm{G}^{-1/2}))\bm{G}^{-1/2})$.}

{2:$\bm{A}^{j+1}=\bm{U}\bm{V}^{H}$.}

{end for }

{3: Use one reference symbol to eliminate the phase
ambiguity by (\ref{eq:phase}) and obtain $\tilde{\bm{X}}$.}

{4: Use the $\lceil\log_{|\mathcal{S}|}K\rceil$ user
ID to eliminate the permutation ambiguity and obtain $\bm{\hat{X}}$.}
\end{algorithm}

\subsection{{Preconditioning for Small Frame Length $T$}}

{For the scenarios where $T$ is small, the data matrix
$\bm{X}$ may not be well concentrated on the Stiefel manifold,
according to Proposition \ref{prop:Signal-concentration}. However,
an efficient preconditioning, as summarized below, can be adopted
to address this issue:}
\begin{itemize}
\item {Step 1. Calculate $\bar{\bm{Y}}_{pre}=\bm{U}_{Y}\bm{V}_{Y}^{H}$,
where $[\bm{U}_{Y},\sim,\bm{V}_{Y}]=SVD_{compact}(\bar{\bm{Y}})$.}
\item {Step 2. Use $\bar{\bm{Y}}_{pre}$ as an input of
Algorithm \ref{alg2} and obtain $\bm{\hat{X}}_{pre}$. }
\item {Step 3. The final detection result is $\bm{\hat{X}}=(\bm{D}^{H}\bm{D})^{-1}\bm{D}^{H}\bar{\bm{Y}}$,
where $\bm{D}=\bar{\bm{Y}}_{pre}\bm{\hat{X}}_{pre}^{H}$.}
\end{itemize}
{In the following, we present a brief interpretation.
When $M$ is large enough, }{\emph{with high probability}}{
we have \cite[Section F]{sun2016complete2}
\begin{equation}
\bar{\bm{Y}}_{pre}=c\bar{\bm{H}}\bm{G}^{1/2}\bm{U}_{X}\bm{V}_{X}^{H}+\bar{\bm{H}}\text{\ensuremath{\boldsymbol{\Delta}}},
\end{equation}
 where $\text{\ensuremath{\boldsymbol{\Delta}}}$ is an error matrix
with small magnitude, and $\bm{U}_{X}\Sigma\bm{V}_{X}^{H}=SVD_{compact}(\bm{X}_{ture})$,
$c$ is a constant. Though $\bm{X}_{ture}$ is not concentrated on
the Stiefel manifold when $T$ is small, $\bm{U}_{X}\bm{V}_{X}^{H}$
satisfies $(\bm{U}_{X}\bm{V}_{X}^{H})^{H}(\bm{U}_{X}\bm{V}_{X}^{H})=\bm{I}$.
Hence $\bm{U}_{X}\bm{V}_{X}^{H}$ is on the Stiefel manifold. Therefore,
using $\bar{\bm{Y}}_{pre}$ as the input of Algorithm \ref{alg2},
we can obtain $\bm{\hat{X}}_{pre}$, which is an estimate of $(\bm{U}_{X}\bm{V}_{X}^{H})^{H}$.
Then, $\bm{D}=\bar{\bm{Y}}_{pre}\bm{\hat{X}}_{pre}^{H}$ can be
considered as an estimate of $c\bm{\bar{\bm{H}}}\bm{G}^{1/2}$.
According to Eq. (\ref{eq:sigsparse}), the estimate of $\bm{X}_{ture}$
can be obtained by $\bm{\hat{X}}=(\bm{D}^{H}\bm{D})^{-1}\bm{D}^{H}\bar{\bm{Y}}$
after row normalization which eliminates the influence of $c$.}

\section{Simulation results\label{sec:Simulation-results}}

In this section, we first numerically verify the convergence property
of the algorithm proposed in Section \ref{sec:Proposed-Parameter-free-Algorith},
and then present comprehensive simulation results to show the superiority
of the proposed scheme for blind data detection in massive MIMO systems.

\subsection{Convergence Property}

\begin{figure}
\centering{}\includegraphics[scale=0.65]{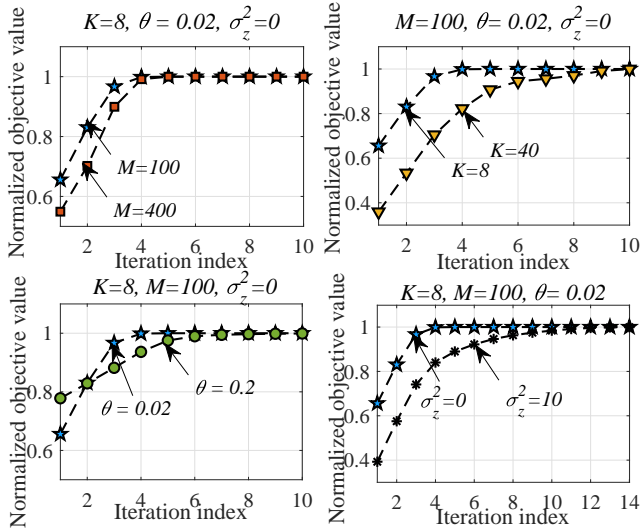}\caption{\label{fig:convespeed}Evaluation of the influence of different system
parameters on the convergence. The convergence trajectory of one random
trial of using Algorithm \ref{alg2} to solve Problem (\ref{eq:Pl3}).
$\bm{X}$ and $\bar{\bm{H}}$ are randomly generated such
that $\bm{X}^{H}\in St_{K}(\mathbb{C}^{T})$ and $\bar{H}_{m,k}\sim_{i.i.d}\mathcal{BG}(\theta)$
with $T=200$. }
\end{figure}

The convergence property is verified with randomly generated $\bm{X}$
and $\bar{\bm{H}}$, such that\footnote{$\bm{X}^{H}\in St_{K}(\mathbb{C}^{T})$ can be generated via the
QR decomposition of any random matrix.} $\bm{X}^{H}\in St_{K}(\mathbb{C}^{T})$ and $\bar{H}_{m,k}\sim_{i.i.d}\mathcal{BG}(\theta)$.
Without loss of generality, we fix $\bm{G}=\bm{I}$ in the
following results.%
{} Fig. \ref{fig:convespeed} considers the noisy case, i.e., $\bm{P}=\bm{I}$
and $\bar{\bm{Z}}$ is the additive Gaussian channel noise with
zero mean and variance $\sigma_{z}^{2}$, to illustrate how the key
system parameters influence the convergence rate. The value of the
objective function is normalized by $\frac{3}{4}\sqrt{\text{\ensuremath{\pi}}}MK\Big(\theta(\sigma_{z}^{2}+1)^{\frac{3}{2}}-\theta(\sigma_{z}^{2})^{\frac{3}{2}}+(\sigma_{z}^{2})^{\frac{3}{2}}\Big)$,
which is the theoretical maximum value of the objective function with
a sufficiently large $M$ (further details about this value can be
found in Appendix \ref{subsec:Proof-of-Proposition-1}). The curves
are plotted for individual trials with different experimental settings
when $T=200$. The results imply that the convergence rate is influenced
by $\theta$, $K$, and the noise variance $\sigma_{z}^{2}$ . Specifically,
a smaller $M$, a smaller $K$, a smaller $\theta$, or a smaller
noise variance $\sigma_{z}^{2}$ will accelerate the convergence rate
consistent with Theorem \ref{thm:(Convergence-of-st-FW)-1}.

\subsection{Performance Evaluation}

In this section, we evaluate the proposed method, in comparison with
existing ones. In the following results, the spatial channel matrix
$\bm{H}$ is generated according to Eq. (\ref{eq:channelgen})
with $N_{l}(k)=5$ paths and i.i.d. Gaussian $\alpha_{ilk}$ with
unit variance for all $K$ users. Each user has independent azimuth
and elevation AoAs, $\varphi_{lk}^{r}$ and $\theta_{lk}^{r}$, which
are assumed to be uniformly distributed in $[0,2\pi)$ and $[-\frac{\pi}{2},\frac{\pi}{2})$.
The antenna elements in the URPA are separated by a half-wavelength
distance. The ``virtual angular domain'' channel $\bm{\bm{\bar{\bm{H}}}}$
is obtained according to Eq. (\ref{eq:angular}). Each element of
$\bm{X}$ is drawn from the i.i.d. QPSK symbols and normalized
by $\frac{1}{\sqrt{T}}$. All the results are conducted by averaging
over $100$ Monte Carlo trials, unless otherwise specified.

\subsubsection{Performance and complexity comparison with baselines}

To demonstrate the benefits of the proposed method, we introduce the
following four state-of-the-art blind data detection schemes as baselines.
Note that we apply the ambiguity elimination scheme described in Section
\ref{subsec:Ambiguity-mitigation} to all the approaches.
\begin{itemize}
\item \textbf{Baseline 1 }{(Blind Pro-Bi-GAMP):
}\cite{zhang2017blind}\textbf{:} This is an approximate probabilistic
message passing algorithm for blind data detection. In the simulation,
we use EMBiGAMP\_DL \cite{Vila:TSP:13} in the GAMP Matlab package\footnote{The Matlab code can be downloaded at https://sourceforge.net/projects/gampmatlab/.}
and add the projection operation illustrated in\cite{zhang2017blind}.
This baseline is introduced to show the robustness of the proposed
method to different distributions of the channel and data.
\item \textbf{Baseline 2}{{{} (Blind $\ell_{1}$):}}\textbf{
}For $\ell_{1}$-norm-based methods, we adopt the one proposed in
\cite{bai2018subgradient} for complete (orthogonal) dictionary learning.
In the simulation, we divide the data matrix $\bm{X}\in\mathbb{C}^{K\times T}$
into $T/K$ adjacent squared matrices and implement the $\ell_{1}$-norm-based
method for each squared matrix. This baseline is introduced to show
the effectiveness of the proposed problem formulation.
\item \textbf{Baseline 3 }{{(Blind $\ell_{4}$):
}}This baseline solves the $\ell_{4}$-norm-based problem; i.e., $||\cdot||_{3}^{3}$
in Problem (\ref{eq:Pl3}) is changed into $||\cdot||_{4}^{4}$. A
similar scheme has been used in image processing to solve an orthogonal
dictionary learning problem\cite{Underl4} in the real-value domain.
This baseline is introduced to show the importance of the choice of
$p$ for the high-order-norm-based problem formulation.
\item \textbf{Baseline 4 }{{(Blind GD $\ell_{3}$):}}\textcolor{blue}{{}
}The gradient algorithm over the Stiefel manifold \cite{li2019nonsmooth}
is used to solve the proposed Problem (\ref{eq:Pl3}). This baseline
is introduced to show the efficiency of the proposed algorithm.
\end{itemize}
We use the average error vector magnitude (EVM) as the performance
metric for data detection. The specific expression is given by

\[
\text{EVM}=\frac{1}{K}\sum_{k=1}^{K}\frac{||\hat{\bm{X}}_{k,:}-\bm{X}_{k,:}||_{2}^{2}}{||\bm{X}_{k,:}||_{2}^{2}}.
\]
We also compare the proposed scheme with the following training-based
sparsity-exploiting data detection scheme.
\begin{itemize}
\item \textbf{Baseline 5 }{{(Pilot-based):}}{
\cite{destino2015leveraging}: }This baseline is a pilot-based method
which leverages the sparsity of channel. In the channel estimation
phase, randomly generated training symbols $\bm{X}_{T}$ with
length $T_{t}$ are sent to the BS. The sparse channel is then estimated
by solving the popular regularized least-squares problem: $\underset{\bar{\bm{H}}}{\text{minimize}}\quad||\bar{\bm{Y}}_{T}-\bm{\bar{\bm{H}}}\bm{G}^{1/2}\bm{X}_{T}||_{2}^{2}+\lambda||\bar{\bm{H}}||_{1}$,
where $\bar{\bm{Y}}_{T}$ is the received signal in the training
period. After $\bar{\bm{H}}$ is estimated, the transmitted data
is detected via zero-forcing (ZF). {The problem is
solved by the alternating direction method of multipliers (ADMM) algorithm
\cite{ADMM} with $\lambda=2$.} 
\end{itemize}
To facilitate a comprehensive comparison, we introduce
the following achievable rate metrics

\[
\text{R}_{\text{{blind}}}=\sum_{k=1}^{K}(1-\frac{1}{T})\log_{2}\Big(1+\frac{||\bm{X}_{k,:}||_{2}^{2}}{||\hat{\bm{X}}_{k,:}-\bm{X}_{k,:}||_{2}^{2}}\Big)-\frac{K\left\lceil \log_{2}K\right\rceil }{T},
\]

\[
\text{R}_{\text{{training}}}=\sum_{k=1}^{K}(1-\frac{T_{t}}{T})\log_{2}\Big(1+\frac{||\bm{X}_{k,:}||_{2}^{2}}{||\hat{\bm{X}}_{k,:}-\bm{X}_{k,:}||_{2}^{2}}\Big),
\]
for the blind and training-based schemes, respectively. For the blind
schemes, the overhead $(1-\frac{1}{T})$ is caused by the common reference
symbol to eliminate the phase ambiguity, and the loss $\frac{K\left\lceil \log_{2}K\right\rceil }{T}$
is caused by the permutation ambiguity. We compare the performance
of the proposed method with the aforementioned baselines in terms
of the EVM, computation time and achievable rate with $N_{h}=M$,
$N_{v}=1$, where $\bm{G}=\bm{I}$,$\bm{P}=P\bm{I}$,
and $\bm{Z}$ is generated as the white Gaussian noise with variance
$\sigma_{z}^{2}=\frac{K}{\text{SNR}T}$. 

\begin{figure}[!]
\centering{}%
\includegraphics[scale=0.5]{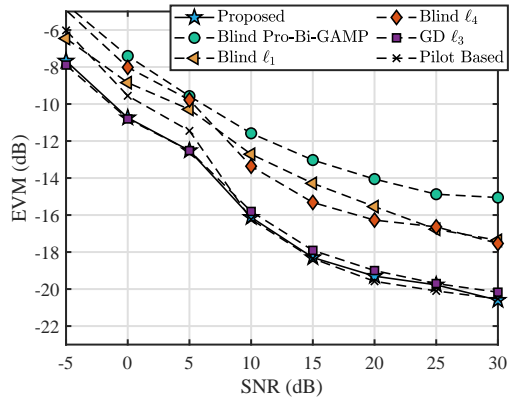}
{\caption{\label{fig:realcom}{The EVM performance comparison
with $N_{h}N_{v}=256$, $T=240$ and $K=8$ under different SNR values.}}}
\end{figure}

Fig. \ref{fig:realcom} shows the performance comparison under different
SNR, with $N_{h}N_{v}=256$, $T=240$ and $K=8$. We see that the proposed
scheme\footnote{The rounding technique in \cite{qu2019nonconvex} is adopted for the
$\ell_{3}$-based method.} exhibits the best performance among the blind schemes. The proposed
scheme outperforms Baselines 1 and 2 because that it is more robust
to noise and the power leakage phenomena. Specifically, the sparsity
penalty $||\cdot||_{3}^{3}$ is a milder sparsity penalty which is
very flat around 0 and insensitive to noise in the signal, while the
strict sparsity penalties used in Baselines 1 and 2 essentially encourage
all small entries to be 0 \cite{zhang2019structured}. Baseline 3
shows inferior performance since it requires a higher sample complexity
than the proposed $\ell_{3}$-norm-based formulation \cite{shen2020complete}.
{Although Baseline 4 shows a competitive performance to the proposed
one, the complexity comparison given in Table \ref{tab:Complexity-Comparison-reak}
indicates its inefficiency compared to the proposed method. Moreover,
the proposed blind scheme is more robust than the pilot-based approach
(Baseline 5) with six pilot symbols in the low SNR region. It is also
comparable with the pilot-based approach in the high SNR region, while
avoiding transmitting pilots for channel estimation.}

\begin{figure}[!]
\centering{}%
\includegraphics[scale=0.5]{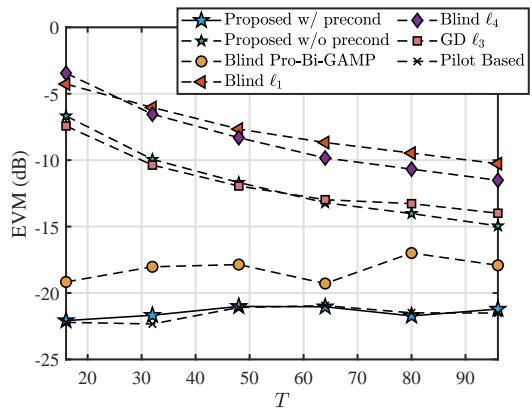}
{\caption{\label{fig:diffT}{The EVM performance comparison
with $N_{h}N_{v}=256$, SNR$=30$dB and $K=8$ under small $T$.}}}
\end{figure}

{Fig. \ref{fig:diffT} shows the performance comparison
under small $T$, with $N_{h}N_{v}=256$, SNR$=30$dB and $K=8$. We see that the proposed
scheme with preconditioning exhibits the best performance in terms
of the EVM among all the blind schemes, while achieving a similar
performance as the pilot-aided method, Baseline 5, and avoiding the
pilot overhead for channel estimation.}

{The complexity comparison in Table \ref{tab:Complexity-Comparison-reak}
is for $\text{SNR}=30\text{dB}$, with other settings the same as
those in Fig. \ref{fig:realcom}. $I$ and $J$ represent the number
of inner iterations and outer iterations, respectively. The results
indicate that the proposed method achieves the second lowest complexity
considering the per-iteration complexity and the number of iterations.
Nevertheless, compared with the one with the lowest complexity, i.e.,
the $\ell_{4}$-based approach, the proposed method achieves much
better performance as shown in Fig. \ref{fig:realcom}. Hence the
proposed scheme is a promising solution to achieve the best performance
and complexity trade off. }

\begin{table*}[!h]
\centering{}{\caption{\label{tab:Complexity-Comparison-reak}{Complexity
Comparison}}
}{\footnotesize{}}%
\begin{tabular}{cccc}
\toprule 
{\footnotesize{}Method} & {\footnotesize{}Number of FLOPs} & {\footnotesize{}I} & {\footnotesize{}J}\tabularnewline
\midrule
\midrule 
{\footnotesize{}Blind Pro-Bi-GAMP} & {\footnotesize{}$\mathcal{O}(JIMTK)$\cite{kuai2020double}} & {\footnotesize{}1500} & {\footnotesize{}21}\tabularnewline
\midrule 
{\footnotesize{}Blind $\ell_{1}$} & {\footnotesize{}$\mathcal{O}((J(MK^{2}+MK+K)+MK^{2}+K^{3})\frac{T}{K})$} & {\footnotesize{}N/A} & {\footnotesize{}4000}\tabularnewline
\midrule 
{\footnotesize{}Blind $\ell_{4}$} & {\footnotesize{}$\mathcal{O}(J\cdot(MTK+MK+TK^{2}))$} & {\footnotesize{}N/A} & {\footnotesize{}19}\tabularnewline
\midrule 
{\footnotesize{}Blind GD $\ell_{3}$} & {\footnotesize{}$\mathcal{O}(J\cdot(MTK+MK+T^{2}K+I(K^{3}+TK^{2}+MTK)))$} & {\footnotesize{}168} & {\footnotesize{}87}\tabularnewline
\midrule 
{\footnotesize{}Pilot-based} & {\footnotesize{}$\mathcal{O}(MJ\cdot(K^{3}+K^{2}T_{t}+K^{2}+KT_{t})+K^{3}+K^{2}M+K^{2}T+KMT)$\cite{ADMM}} & {\footnotesize{}N/A} & {\footnotesize{}39}\tabularnewline
\midrule 
{\footnotesize{}Proposed method} & {\footnotesize{}$\mathcal{O}(J\cdot(MTK+MK+TK^{2}))$} & {\footnotesize{}N/A} & {\footnotesize{}26}\tabularnewline
\bottomrule
\end{tabular}
\end{table*}

\begin{figure}[H]
\centering{}%
\noindent\begin{minipage}[t]{1\columnwidth}%
\begin{center}
\includegraphics[scale=0.35]{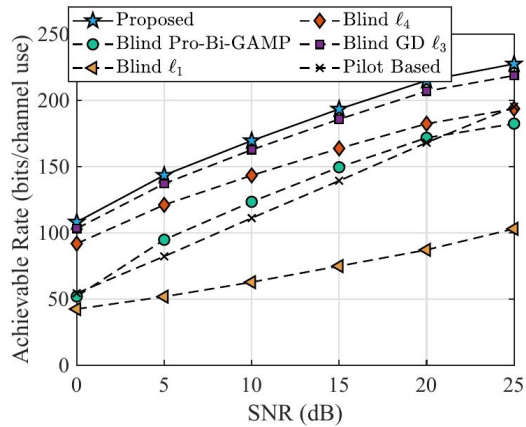}
\par\end{center}
\caption{\label{fig:mmwavec} The average achievable rate performance comparison.}
\end{minipage}
\end{figure}

We also compare the proposed method with other baselines in terms
of the achievable rate \footnote{Preconditioning is adopted for all methods that leverage the data concentration.} with $N_{h}N_{v}=784$, $T=120$, and $K=30$
in Fig. \ref{fig:mmwavec}. As we see, the proposed method outperforms
the other baselines including the pilot-based method (Baseline 5)
with $20$ pilot symbols, which suffers a rate loss caused by the
pilot overhead and estimation errors.

\subsubsection{Impact of key system parameters}

\begin{figure*}
	\centering
	\begin{subfigure}{.3\textwidth}
		\centering
		\includegraphics[width=.95\linewidth]{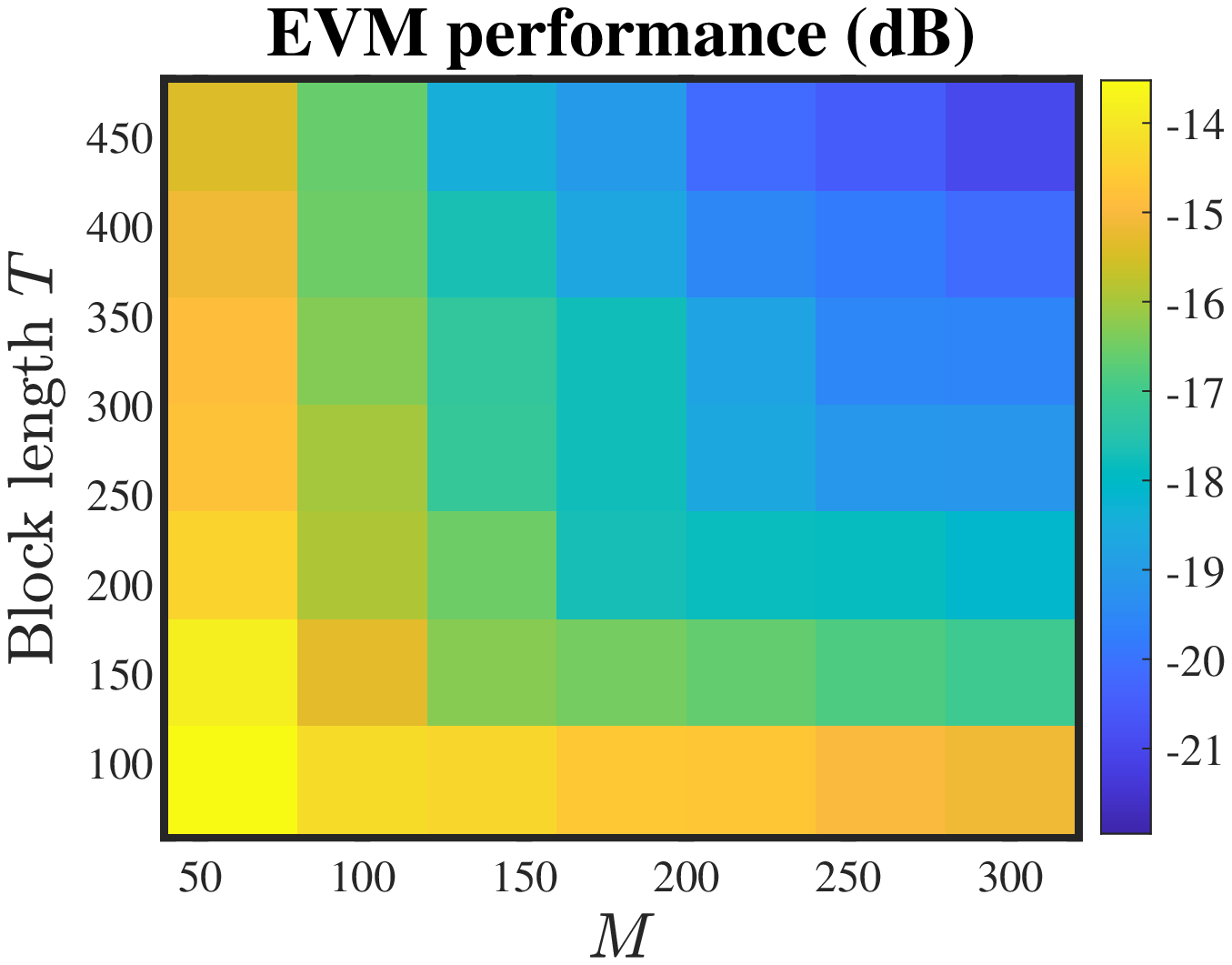}
		\caption{Average EVM performance under
			different numbers of $M$ and $T$ with $\text{SNR}=20\text{dB}$.}
		\label{fig:realdiffT-1}
	\end{subfigure}%
~
	\begin{subfigure}{.3\textwidth}
		\centering
		\includegraphics[width=.95\linewidth]{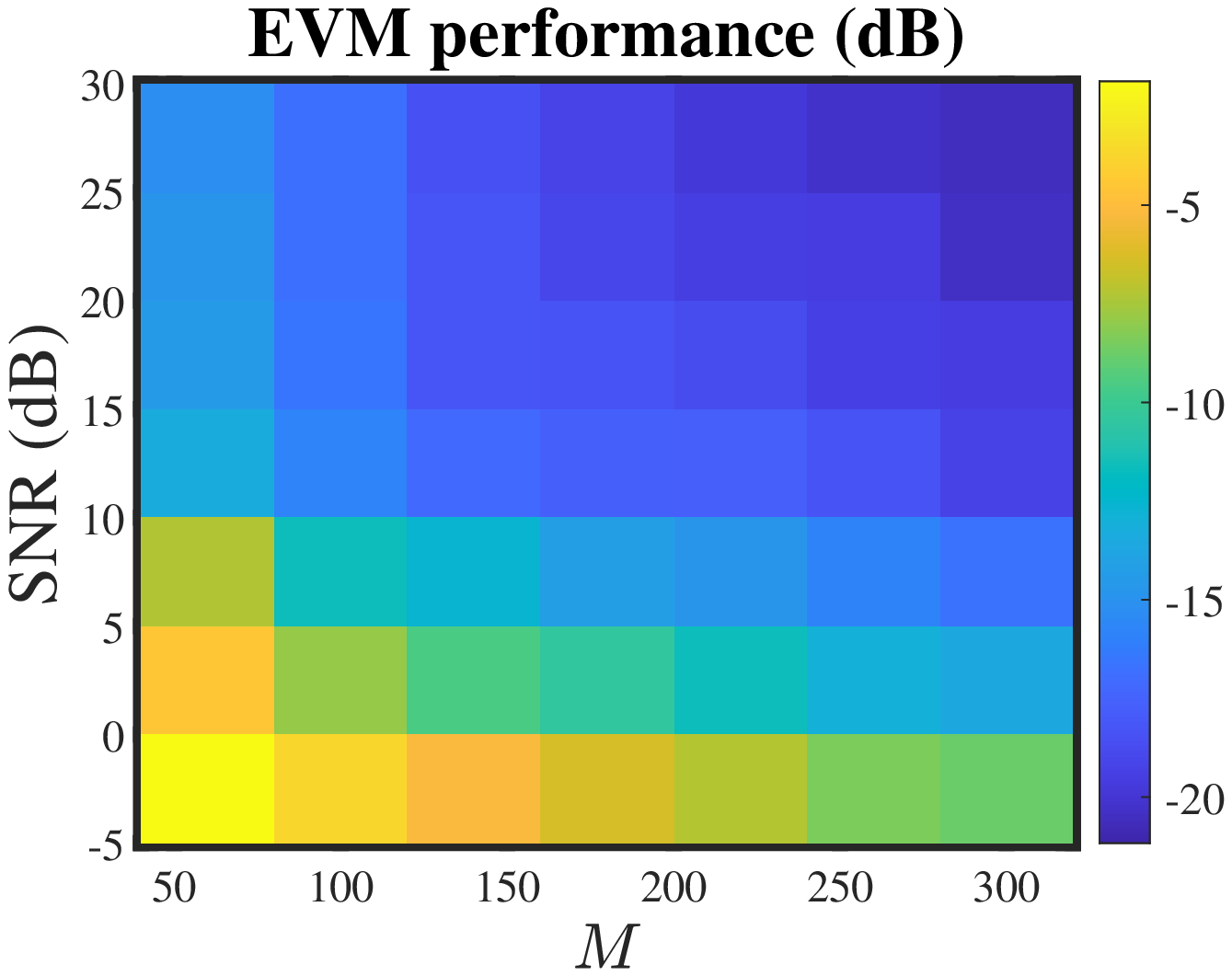}
		\caption{Average EVM performance under different
			numbers of $M$ and SNR with $T=400$.}
		\label{fig:realdiffSNR-1}
	\end{subfigure}
~
\begin{subfigure}{.3\textwidth}
	\centering
	\includegraphics[width=.95\linewidth]{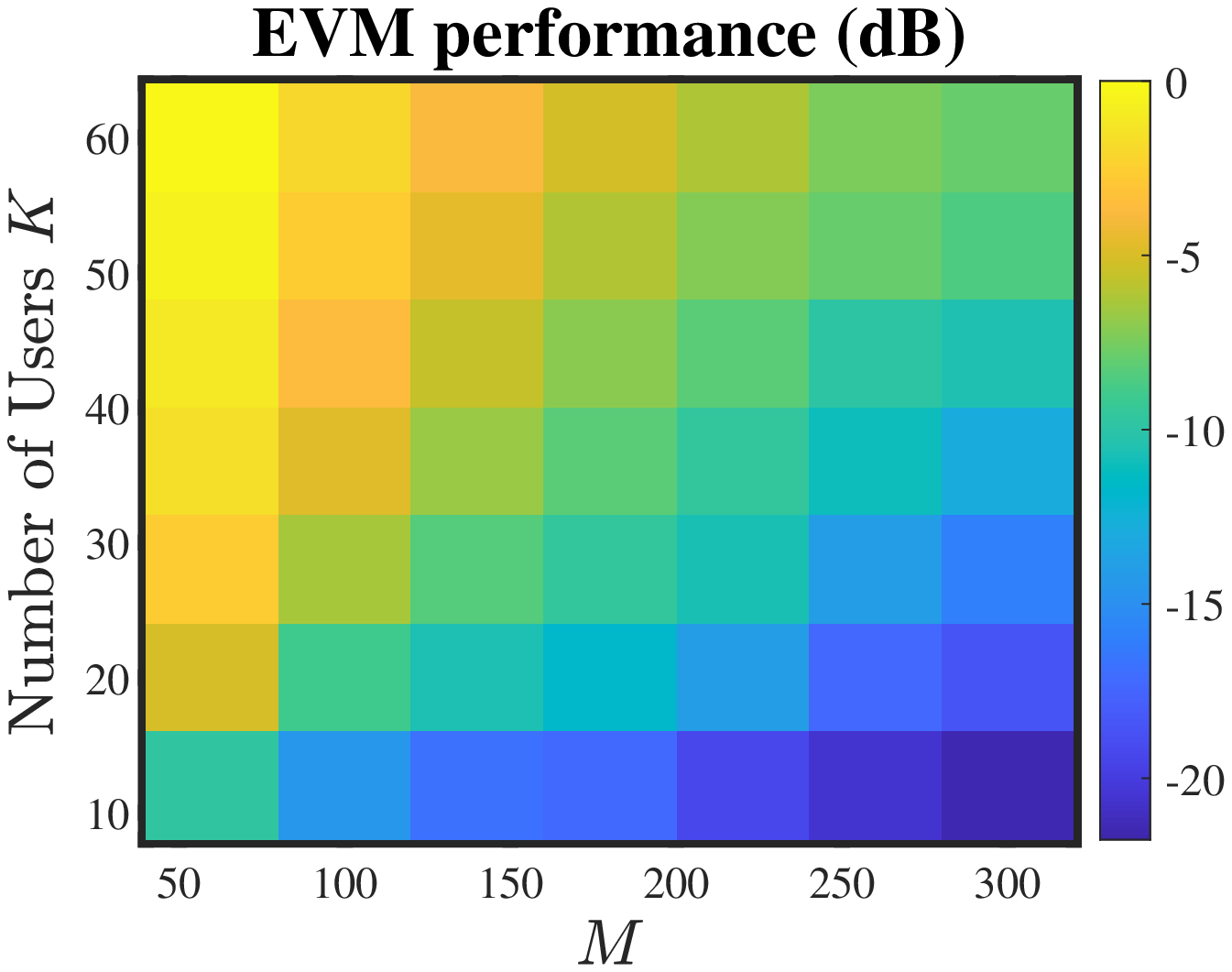}
	\caption{Average EVM performance under
		different numbers of $M$ and $K$ with $T=400$, SNR = $20$dB.}
	\label{fig:realdiffK}
\end{subfigure}
	\caption{Average EVM performance under different key system parameters.}
	\label{fig:Impact-of-key-1}
\end{figure*}

We test the impact of key system parameters on the performance of
the proposed scheme with $N_{h}=M$ and $N_{v}=1$, averaging over
$2000$ Monte Carlo trials. We set $\bm{G}=\bm{I}$, $\bm{P}=\bm{I}$,
$\bm{Z}$ with variance $\sigma_{z}^{2}=\frac{K}{\text{SNR}T}$
and $K=8$. Fig. \ref{fig:realdiffT-1} shows the EVM performance
with $\text{SNR}=20$ dB. We see that a larger $T$ leads to a better
performance since $\bm{X}^{H}$ concentrates closer to the Stiefel
manifold as $T$ increases, according to Proposition \ref{prop:Signal-concentration}.
The influence of the SNR with $T=400$ is given in Fig. \ref{fig:realdiffSNR-1}.
As it can be seen, for a fixed number of $M$, a higher SNR leads
to better performance and a large $M$ makes recovery in the low SNR
region possible. These observations verify the results in Theorem
\ref{thm:Asymptotically-exact-recovery-1}. {The influence
of the number of users $K$ is given in Fig. \ref{fig:realdiffK}.
We see that a smaller $K$ leads to better performance which is consistent
with Theorem \ref{thm:Asymptotically-exact-recovery-1}.}

\subsubsection{EVM performance with different large-scale fading}

\begin{figure}[htpb]
\centering{}%
\noindent\begin{minipage}[H]{1\columnwidth}%
\begin{center}
\includegraphics[scale=0.5]{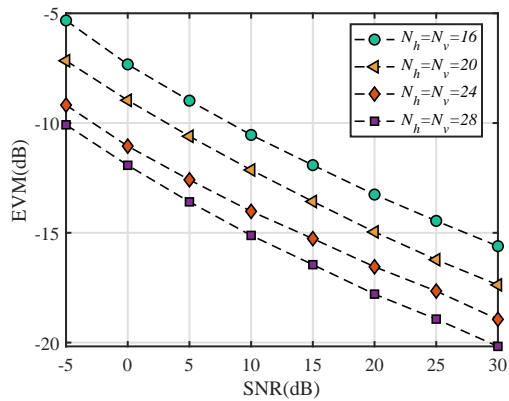}
\par\end{center}
\caption{\label{fig:realcomlarge}The EVM performance of the proposed scheme
considering different values of large-scale fading of each user.}
\end{minipage}
\end{figure}

Fig.\ref{fig:realcomlarge} shows the EVM performance with $N_{h}=N_{v}=\sqrt{M}$
and different large-scale fading among $K$ users. The large-scale
fading of the $k$-th user is generated by $G_{k,k}=-32.4-18.5log_{10}(d_{k})-20log_{10}(f_{c})+\text{\ensuremath{\chi_{\sigma_{Sf}}}}$
(dB) according to \cite[Table 4]{rappaport2017overview}, where $d_{k}$
is the 3D distance between the $k$-th user and the BS, which is randomly
drawn from $(20,200)$, $f_{c}=28$ GHz, and $\text{\ensuremath{\chi_{\sigma_{Sf}}}}\sim\mathcal{CN}(0,4.2)$.
$\bm{P}=\bm{I}$ and $\bm{Z}$ is generated as the white
Gaussian noise with variance $\sigma_{z}^{2}=\frac{\sum_{k=1}^{K}G_{k,k}}{T\text{SNR}}$,
$T=240$ and $K=8$. Though the power leakage is more severe when
$N_{h}=N_{v}=\sqrt{M}$ , the results show the robustness of the proposed
scheme with large receive antenna arrays. 

\section{Conclusions\label{sec:Conclusion}}

In this paper, we proposed to exploit the sparsity of the massive MIMO channel and the data concentration for blind data detection. An $\ell_{3}$-norm maximization problem 
was proposed, along with theoretical justification. A parameter-free
algorithm was then proposed to solve the problem with a convergence
guarantee. Numerical-simulation-based results were provided to verify
the correctness of the derived theorems, as well as, to demonstrate
the superior performance and robustness of the proposed blind data
detection method.  {It will be useful to
develop a stochastic version for the proposed scheme  to further reduce
the complexity and facilitate the implementation in the large-scale
OFDM mmWave systems.}

\appendix

\subsection{\label{subsec:Proof-of-Proposition}Proof of Proposition \ref{prop:Signal-concentration}}

Since any bounded random variable $X$ is sub-Gaussian with $||X||_{\psi_{2}}\leq\frac{1}{\sqrt{\ln2}}||X||_{\infty}$,
where $||\cdot||_{\psi_{2}}$ is the sub-Gaussian norm defined as
$||X||_{\psi_{2}}=\inf\{\sigma>0:\mathbb{E}[e^{(|X|^{2}/\sigma^{2})}\leq2]\}$
\cite{vershynin2018high}, and $\sqrt{T}\bm{X}\in\mathbb{R}^{K\times T}$
has independent, mean-zero, sub-Gaussian isotropic random columns,
i.e., $\mathbb{E}[\bm{X}{}_{:,t}(\bm{X}_{:,t})^{H}]=\bm{I}_{K},\forall t=1,\ldots,T$.
Then, for any $\delta>0$, we obtain
\begin{equation}
\begin{aligned}
&Pr\Big[\frac{||\bm{X}\bm{X}^{H}-\bm{I}||_{F}}{\sqrt{K}}\leq\frac{1}{\ln2}\mathcal{S}_{\infty}^{2}\max\{C\sqrt{\frac{K}{T}}+\frac{\delta^{\prime}}{\sqrt{T}},\\
&(C\sqrt{\frac{K}{T}}+\frac{\delta^{\prime}}{\sqrt{T}})^{2}\}\Big]\\
\geq & Pr\Big[||\bm{X}\bm{X}^{H}-\bm{I}||\leq\frac{1}{\ln2}\mathcal{S}_{\infty}^{2}\max\{C\sqrt{\frac{K}{T}}+\frac{\delta^{\prime}}{\sqrt{T}},\\
&(C\sqrt{\frac{K}{T}}+\frac{\delta^{\prime}}{\sqrt{T}})^{2}\}\Big]\\
= & Pr\Big[||\frac{1}{T}\sqrt{T}\bm{X}\sqrt{T}(\bm{X})^{T}-\bm{I}||\leq\frac{1}{\ln2}\mathcal{S}_{\infty}^{2}\\
&\max\{C(\sqrt{\frac{K}{T}}+\frac{\delta^{\prime}}{\sqrt{T}}),C^{2}(\sqrt{\frac{K}{T}}+\frac{\delta^{\prime}}{\sqrt{T}})^{2}\}\Big]\\
\ensuremath{\geq} & 1-2\text{\ensuremath{\exp(}}-\delta^{\prime}{}^{2}),
\end{aligned}
\end{equation}
where $\delta^{\prime}\geq0$. Let $\delta=C(\sqrt{\frac{K}{T}}+\frac{\delta^{\prime}}{\sqrt{T}})$,
and then $\delta^{\prime}=\frac{\delta\sqrt{T}}{C}-\sqrt{K}$. Since
$\delta^{\prime}\geq0$, we have $T\geq\frac{C^{2}K}{\delta^{2}}$.
The last inequality follows \cite[Theorem 4.6.1]{vershynin2018high}.

\subsection{\label{subsec:Proof-of-Theorem}Proof of Theorem \ref{thm:Asymptotically-exact-recovery}}

Let $\bm{W}=\bm{X}_{ture}\bm{A}\in\mathbb{C}^{K\times K}$,
and we have
\begin{equation}
\begin{aligned}
&\mathbb{E}[||\bar{\bm{Y}}\bm{A}||_{3}^{3}  =\sum_{m=1}^{m=M}\mathbb{E}[||\bar{\bm{H}}_{m,:}\bm{W}||_{3}^{3}]=M\sum_{k=1}^{k=K}\mathbb{E}[|\bar{\bm{H}}_{m,:}\bm{W}_{:,k}|^{3}]\\
=&M\sum_{k=1}^{k=K}\mathbb{E}[|\langle\bm{W}_{:,k}\odot\bm{b},\bm{g}\rangle|^{3}]\\
\end{aligned}
\end{equation}
 where we denote $\bm{b}\sim_{i.i.d}\mathcal{B}(\theta)$ and
$\bm{g}\sim_{i.i.d}\mathcal{CN}(0,1)$. Using the rotation-invariant
property of Guassian random variables, we have $\ensuremath{\mathbb{E}[|\langle\bm{W}_{:,k}\odot\bm{b},\bm{g}\rangle|^{3}]=\gamma_{1}\mathbb{E}||\bm{W}_{:,k}\odot\bm{b}||_{2}^{3}}$
with $\gamma_{1}=\frac{3}{4}\sqrt{\text{\ensuremath{\pi}}}$ calculated
by the 3rd-order non-central moment of the Rayleigh distribution.
Since $||\bm{W}_{:,k}||_{2}=||\bm{X}_{ture}\bm{A}_{:,k}||_{2}\leq1$,
we have $0\leq\mathbb{E}||\bm{W}_{:,k}\odot\bm{b}||_{2}^{3}\leq\mathbb{E}||\bm{W}_{:,k}\odot\bm{b}||_{2}^{2}=\theta$.
The equality holds if and only if $||\bm{W}_{:,k}\odot\bm{b}||_{2}\in\{0,1\}$
for all $\bm{b}$, which is only satisfied at $\bm{W}_{:,k}\in\{e^{j\phi}\bm{e}_{i}:i\in[K],\phi\in\{0,2\pi\}\}$\cite{bai2018subgradient}.
Therefore, we have $\mathbb{E}[||\bar{\bm{Y}}\bm{A}||_{3}^{3}=M\sum_{k=1}^{k=K}\mathbb{E}[||<\bm{W}_{:,k}\odot\bm{b},\bm{g}>||_{3}^{3}]\leq MK\gamma_{1}\theta$.
Furthermore, if $\bm{W}_{:,k_{1}}=e^{j\phi_{1}}\bm{e}_{i}$
and $\bm{W}_{:,k_{2}}=e^{j\phi_{2}}\bm{e}_{i}$, then $Tr(\bm{W}_{:,k_{1}}^{H}\bm{W}_{:,k_{2}})=e^{j(\phi_{1}+\phi_{2})}$.
However, $Tr(\bm{W}_{:,k_{1}}^{H}\bm{W}_{:,k_{2}})=Tr((\bm{X}_{ture}\bm{A}_{:,k_{1}})^{H}\bm{X}_{ture}\bm{A}_{:,k_{2}})=Tr(\bm{A}_{:,k_{1}}^{H}\bm{A}_{:,k_{2}})=0$.
This indicates that two different columns of $\bm{W}$ cannot
simultaneously equal the same standard basis vector. Hence,  $\mathbb{E}[||\bar{\bm{Y}}\bm{A}||_{3}^{3}$
achieves the maximum $MK\gamma_{1}\theta$ when $\bm{A}=\bm{A}^{opt}$
with $\bm{X}_{ture}\bm{A}^{opt}=\boldsymbol{\Xi}^{H}$. 

Using the result in \cite[Theorem 2.1]{shen2020complete} for the
Stiefel manifold, we conclude that there exists a constant $c\geq0$,
for any $\delta>0$, such that, whenever$M\geq c\theta\delta^{-2}T\log(K/\delta)(K\log^{2}K)^{\frac{3}{2}}$,
$Pr\Big(\frac{1}{K}||\bm{A}^{opt}\boldsymbol{\Xi}-\bm{X}_{ture}^{H}||_{F}^{2}\leq\delta\Big)\geq1-M^{-1}$. 

\subsection{\label{subsec:Proof-of-Proposition-1}Proof of Proposition \ref{thm:Asymptotically-exact-recovery-1}}

Let $\bm{W}=\bm{\bm{X}}\bm{A}$, and then we have
\begin{equation}
\begin{aligned}
&\mathbb{E}_{\bm{\bar{\bm{H}}},\bar{\bm{Z}}}[||\bar{\bm{Y}}\bm{\bm{A}}\bm{G}^{-1/2}||_{3}^{3}] \\ =&\sum_{m=1}^{m=M}\mathbb{E}[||\bar{\bm{H}}_{m,:}\bm{G}^{1/2}\bm{W}\bm{G}^{-1/2}+\bar{\bm{Z}}_{m,:}\bm{A}\bm{G}^{-1/2}||_{3}^{3}]\\
 =&M\sum_{k=1}^{k=K}\mathbb{E}_{\bm{b},\bm{g}}[|\langle[\bm{W}_{:,k}\odot\bm{b};\sigma_{z}G_{k,k}^{-1/2}\bm{A}_{:,k}],\bm{g}\rangle|^{3}],
\end{aligned}
\end{equation}
 where we denote $\bm{b}\sim_{i.i.d}\mathcal{B}(\theta)$ and
$\bm{g}\sim_{i.i.d}\mathcal{CN}(0,1)$. Using the rotation invariant
property of Gaussian random variables, we have $\ensuremath{\mathbb{E}_{\bm{b},\bm{g}}[|\langle[\bm{W}_{:,k}\odot\bm{b};\sigma_{z}G_{k,k}^{-1/2}\bm{A}_{:,k}],\bm{g}\rangle|^{3}]=\gamma_{1}\mathbb{E}(||\bm{W}_{:,k}\odot\bm{b}||_{2}^{2}+(G_{k,k}/\sigma_{z}^{2})^{-1}})^{\frac{3}{2}}$
with $\ensuremath{\gamma_{1}}=\frac{3}{4}\sqrt{\pi}$ calculated by
the third-order non-central moment of the Rayleigh distribution. Then,
we have 
\begin{equation}
\begin{aligned}
&\gamma_{1}\theta M\sum_{k=1}^{K}((G_{k,k}/\sigma_{z}^{2})^{-1})^{\frac{3}{2}}\\ \leq&\mathbb{E}_{\bm{\bar{\bm{H}}},\bar{\bm{Z}}}[||\bar{\bm{Y}}\bm{\bm{A}}\bm{G}^{-1}||_{3}^{3}]\\
\leq & \gamma_{1}M(\sum_{k=1}^{K}\theta\Big((1+(G_{k,k}/\sigma_{z}^{2})^{-1})^{\frac{3}{2}}-((G_{k,k}/\sigma_{z}^{2})^{-1})^{\frac{3}{2}}\Big)\\
&+((G_{k,k}/\sigma_{z}^{2})^{-1})^{\frac{3}{2}}).
\end{aligned}
\end{equation}

The first equality holds when $\bm{W}_{:,k}=\bm{0}$ and the
second equality holds when $\bm{W}_{:,k}\in\{e^{j\phi}\bm{e}_{k}:k\in[K],\phi\in[0,2\pi]\}$.

Using the result in \cite[Theorem 2.2]{shen2020complete} for the
Stiefel manifold, we have whenever  $M\geq c\theta\tilde{\delta}^{-2}K\log(K/\tilde{\delta})(\sum_{k=1}^{K}(1+\sigma_{z}^{2}G_{k,k}^{-1})\log^{2}K)^{\frac{3}{2}}\bar{\xi}_{\sigma}^{2}$,
$Pr[\frac{1}{K}||\bm{A}^{opt}\boldsymbol{\Xi}-\bm{X}_{ture}^{H}||_{F}^{2}\leq\tilde{\delta}]\geq1-M^{-1}$
with $\bar{\xi}_{\sigma}=(\sum_{k=1}^{K}((1+(G_{k,k}/\sigma_{z}^{2})^{-1})^{\frac{3}{2}}+((G_{k,k}/\sigma_{z}^{2})^{-1})^{\frac{3}{2}})-\sum_{k=1}^{K}2(0.5+\sigma_{z}^{2}G_{k,k}^{-1})^{\frac{3}{2}})/K^{\frac{3}{2}}>\bar{\xi}$.
Let $\tilde{\delta}=\frac{\delta\xi(\sum_{k=1}^{K}(1+(G_{k,k}/\sigma_{z}^{2})^{-1})\text{)}^{\frac{3}{4}}}{K^{\frac{9}{4}}}$,
we obtain the result in Theorem \ref{thm:Asymptotically-exact-recovery-1}
with $\xi=\bar{\xi}K^{\frac{3}{2}}$.

\subsection{\label{subsec:Proof-of-Lemma-3}Proof of Lemma \ref{lem:If--is}}

Define $\Psi_{n}(\bm{A})=||\bar{\bm{Y}}\bm{A}\bm{G}^{-1/2}||_{3}^{3}$.
Then, by the convexity of $\Psi_{n}(\bm{A})$, we have
\begin{equation}
\begin{aligned}
&\Psi_{n}((1-\upsilon^{j})\bm{A}^{j}+\upsilon^{j}\bm{S}^{j})  \\
\leq&(1-\upsilon^{j})\Psi_{n}(\bm{A}^{j})+\upsilon^{j}\Psi_{n}(\bm{S}^{j}) \\
\leq&\Psi_{n}(\bm{S}^{j})+(\upsilon^{j}-1)\Big(\langle\nabla\Psi_{n}(\bm{A}^{j}),\bm{S}^{j}\rangle-\langle\nabla\Psi_{n}(\bm{A}^{j}),\bm{A}^{j}\rangle\Big)\\
 \leq&\Psi_{n}(\bm{S}^{j}), 
\end{aligned}
\end{equation}
where the first and second inequalities hold due to the convexity,
and the last inequality holds since $\langle\nabla\Psi_{n}(\bm{A}^{j}),\bm{S}^{j}\rangle-\langle\nabla\Psi_{n}(\bm{A}^{j}),\bm{A}^{j}\rangle\geq0$
and $\upsilon^{j}\in(0,1)$, and the equality is obtained when $\upsilon^{j}=1$.

\subsection{\label{subsec: Proof-of-Lemmaopt}Proof of Lemma \ref{lem:(Optimality-Measure)-}}

Define $\Psi_{n}(\bm{A})=||\bar{\bm{Y}}\bm{A}\bm{G}^{-1/2}||_{3}^{3}$.
Then, we have $\langle\bm{A}-\bm{A}^{j},\nabla_{\bm{A}^{j}}\Psi_{n}(\bm{A})\rangle\leq0$,
which indicates that $\bm{A}^{j}=\arg\max_{\bm{A}\in St_{K}(\mathbb{C}^{T})}\langle\nabla_{\bm{A}^{j}}\Psi_{n}(\bm{A}),\bm{A}\rangle=\bm{A}^{j+1}$.
That is to say, when $\eta(\bm{A}^{j})=0$, we have $\bm{A}^{j+1}=\bm{A}^{j}$.
We also have SVD$(\nabla_{\bm{A}^{j}}\Psi_{n}(\bm{A}))=\bm{U}\boldsymbol{\Sigma}\bm{V}^{H}$,
$\bm{A}^{j+1}=\bm{U}\bm{I}_{T\times K}\bm{V}^{H}$,
and the stationary point of $\Psi_{n}(\bm{A}^{j})$ on $St_{K}(\mathbb{C}^{T})$
satisfies \cite{absil2009optimization}

\begin{equation}
\begin{aligned}
&grad\Psi(\bm{A}^{j})=(\bm{I}-\bm{A}^{j}(\bm{A}^{j})^{H})\nabla_{\bm{A}^{j}}\Psi_{n}(\bm{A})\\
&+\frac{1}{2}\bm{A}^{j}((\bm{A}^{j})^{H}\nabla_{\bm{A}^{j}}\Psi_{n}(\bm{A})-\nabla_{\bm{A}^{j}}\Psi_{n}(\bm{A}){}^{H}\bm{A}^{j})=0.
\end{aligned}
\end{equation}
\begin{itemize}
\item If $\eta(\bm{A}^{j})=0$, we have 
\begin{equation}
\begin{aligned}
&grad\Psi(\bm{A}^{j}) =\bm{U\Sigma\bm{V}}^{H}-\bm{U}\bm{I}_{T\times K}\bm{I}_{T\times K}^{H}\Sigma\bm{V}^{H}\\
+&\frac{1}{2}\bm{U}\bm{I}_{T\times K}\bm{I}_{T\times K}^{H}\bm{\Sigma\bm{V}}^{H}-\frac{1}{2}\bm{\bm{U}}\bm{I}_{T\times K}\Sigma^{H}\bm{I}_{T\times K}\bm{V}^{H}\\
& =\bm{U\Sigma\bm{V}}^{H}-\bm{\bm{U}}\bm{I}_{T\times K}\bm{I}_{T\times K}^{H}\Sigma\bm{V}^{H}=0,
\end{aligned}
\end{equation}
where $\bm{I}_{T\times K}=[\bm{I}_{K};\bm{0}_{T-K}]$;
\item If $grad\Psi(\bm{A}^{j})=0$, we have 
\begin{equation}
\begin{aligned}
&\nabla_{\bm{A}^{j}}\Psi_{n}(\bm{A})  =\frac{1}{2}\bm{A}^{j}(\bm{A}^{j})^{H}\nabla_{\bm{A}^{j}}\Psi_{n}(\bm{A})\\
&+\frac{1}{2}\bm{A}^{j}\nabla_{\bm{A}^{j}}\Psi_{n}(\bm{A})^{H}\bm{A}^{j}\\
\Rightarrow & (\bm{A}^{j})^{H}\bm{U}\boldsymbol{\Sigma}\bm{V}^{H}=\bm{V}\boldsymbol{\Sigma}^{H}\bm{U}^{H}\bm{A}^{j}\Rightarrow\bm{A}^{j}=\bm{U}\bm{I}_{T\times K}\bm{V}^{H}\\
\Rightarrow&\eta(\bm{A}^{j})=0.
\end{aligned}
\end{equation}
\end{itemize}

\subsection{\label{subsec:Proof-of-Theorem-dec}Proof of Theorem \ref{lem:(Monotonically-increasing-value}}

Define $\Psi_{n}(\bm{A})=||\bar{\bm{Y}}\bm{A}\bm{G}^{-1/2}||_{3}^{3}$.
By the convexity of $\Psi_{n}(\bm{A})$, we have $\Psi_{n}(\bm{A}^{j+1})\geq\Psi_{n}(\bm{A}^{j})+\langle\nabla_{\bm{A}^{j}}\Psi_{n}(\bm{A}),\bm{A}^{j+1}-\bm{A}^{j}\rangle\geq\Psi_{n}(\bm{A}^{j})$.
By summing these inequalities for $j=0,1,\ldots,$ we obtain $\Psi_{n}(\bm{A}^{opt})-\Psi_{n}(\bm{A}^{0})\geq\Psi_{n}(\bm{A}^{j})-\text{\ensuremath{\Psi_{n}}(\ensuremath{\bm{A}^{0}})}\geq\sum_{i=0}^{j}\eta(\text{\ensuremath{\bm{A}}}^{i})$.
Hence, $\underset{j\to\infty}{\lim}\eta(\text{\ensuremath{\bm{A}}}^{j})=0$
and 
\[
\underset{0\leq i\leq j}{\min}\eta(\text{\ensuremath{\bm{A}}}^{i})\leq\frac{\Psi_{n}(\bm{A}^{opt})-\Psi_{n}(\bm{A}^{0})}{j+1}.
\]

\subsection{\label{subsec:Proof-of-Theorem-noisy}Proof of Theorem \ref{thm:(Convergence-of-st-FW)-1}}

Define $\Psi_{n}(\bm{A})=||\bar{\bm{Y}}\bm{A}\bm{G}^{-1/2}||_{3}^{3}$.
From \cite[Lemma B.8]{shen2020complete}, if the conditions in Theorem
\ref{thm:Asymptotically-exact-recovery-1} hold, there exists a constant
$c\geq0$, for any $\delta>0$, whenever $M\geq c\theta\delta^{-2}T\log(K/\delta)(K\log^{2}K)^{\frac{3}{2}}$.
Then, we have 
\begin{equation}
\begin{aligned}
 & Pr[M\sum_{k=1}^{K}\gamma_{1}((G_{k,k}/\sigma_{z}^{2})^{-1})^{\frac{3}{2}}-\delta\leq\frac{1}{M}||\bar{\bm{Y}}\bm{A}\bm{G}^{-1/2}||_{3}^{3}\\
 \leq& M(\sum_{k=1}^{K}\theta\gamma_{1}\Big(((G_{k,k}/\sigma_{z}^{2})^{-1}+1)^{3/2}-((G_{k,k}/\sigma_{z}^{2})^{-1})^{3/2}\Big)\\
 &+\gamma_{1}((G_{k,k}/\sigma_{z}^{2})^{-1})^{3/2})+\delta]\\
& \geq1-M^{-1}
\end{aligned}
\end{equation}
 Therefore, with high probability, we have $\Psi_{n}(\bm{A}^{0})\geq M\sum_{k=1}^{K}\gamma_{1}((G_{k,k}/\sigma_{z}^{2})^{-1})^{\frac{3}{2}}-\delta$, and
 \begin{equation}
 \begin{aligned}
 &\Psi_{n}(\bm{A}^{opt})\\
 \leq& M(\sum_{k=1}^{K}\theta\gamma_{1}\Big(((G_{k,k}/\sigma_{z}^{2})^{-1}+1)^{3/2}-((G_{k,k}/\sigma_{z}^{2})^{-1})^{3/2}\Big)\\
 &+\gamma_{1}((G_{k,k}/\sigma_{z}^{2})^{-1})^{3/2})+\delta.
 \end{aligned}
 \end{equation}
 Using the result in Theorem \ref{lem:(Monotonically-increasing-value},
we finish the proof.

\section*{Acknowledgment}
The authors would like to thank professor Yi Ma of Berkeley
EECS Department for his lectures and talks at Tsinghua-Berkeley Shenzhen Institute and Yuexiang Zhai of
Berkeley for stimulating discussions during preparation of this manuscript. The authors would also like to thank professor Xiaojun Yuan of UESTC to share the simulation codes and Hang Liu of CUHK for discussion on the ambiguity resolving.

\ifCLASSOPTIONcaptionsoff
\newpage
\fi



%
\bibliographystyle{ieeetr}
\bibliography{IEEEabrv,IEEEexample,IEEEfull,noconvex}
\end{document}